\documentclass[a4paper,11pt]{article}
\pdfoutput=1 
\usepackage{amssymb,mathrsfs}
\usepackage{bbm}
\usepackage{siunitx}
\usepackage{amsmath}
\usepackage{braket}
\usepackage{mathtools}
\usepackage[usenames,dvipsnames,svgnames,table]{xcolor}
\usepackage [utf8] {inputenc}
\usepackage{color}
\usepackage{lmodern}
\usepackage{footnote}
\usepackage{slashed}
 \usepackage{mathrsfs}
 \usepackage[pdftex] {graphicx}
\usepackage{multirow}
\usepackage{jheppub} 
\usepackage{here}
\usepackage{hyperref}
\usepackage[justification=centering, singlelinecheck=false]{caption}
\usepackage[list=true, labelfont=bf, labelformat=brace, position=top]{subcaption}
\newcommand{\eq}{\begin{eqnarray}}
\newcommand{\en}{\end{eqnarray}}

\title{L\"uscher equation with long-range forces}

\renewcommand{\thefootnote}{\fnsymbol{footnote}}

\author[1]{Rishabh Bubna,}
\affiliation[1]{Helmholtz-Institut f\"ur Strahlen- und Kernphysik (Theorie)\\ and Bethe Center for Theoretical Physics, Universit\"at Bonn, 53115 Bonn, Germany}
\emailAdd{bubna@hiskp.uni-bonn.de}

\author[2,3]{Hans-Werner Hammer,}
\affiliation[2]{Technische Universit\"at Darmstadt, Department of Physics,
64289 Darmstadt, Germany}
\affiliation[3]{ExtreMe Matter Institute EMMI and Helmholtz Forschungsakademie
  Hessen f\"ur FAIR (HFHF),
GSI Helmholtzzentrum f\"ur Schwerionenforschung GmbH,
64291 Darmstadt, Germany}
\emailAdd{hans-werner.hammer@physik.tu-darmstadt.de}

\author[1]{Fabian M\"uller,}
\emailAdd{f.mueller@hiskp.uni-bonn.de}

\author[4]{Jin-Yi Pang,\footnote{Corresponding author.}}
\affiliation[4]{College of Science, University of Shanghai for Science and Technology, Shanghai 200093, China}
\emailAdd{jypang@usst.edu.cn}

\author[1,5]{Akaki Rusetsky,}
\affiliation[5]{Tbilisi State  University,  0186 Tbilisi, Georgia}
\emailAdd{rusetsky@hiskp.uni-bonn.de}

\author[6]{and Jia-Jun Wu}
\affiliation[6]{School of Physical Sciences, University of Chinese Academy of Sciences, Beijing 100049, China}
\emailAdd{wujiajun@ucas.ac.cn}

\abstract{

\noindent
  We derive the modified L\"uscher equation in the presence of the long-range force
  caused by the exchange of a light particle. It is shown that the use of this equation
  enables one to circumvent the problems related to the strong partial-wave mixing and
  the $t$-channel sub-threshold singularities.
  It is also demonstrated that the present method is intrinsically
  linked to the so-called modified effective-range expansion (MERE) in the infinite volume.
  A detailed comparison with the two recently proposed alternative approaches is provided.

}

\allowdisplaybreaks
\begin{document}
\maketitle
\flushbottom

\renewcommand{\thefootnote}{\arabic{footnote}}

\section{Introduction}

In recent decades,
the L\"uscher method~\cite{Luscher:1990ux} has become a standard tool for the
extraction of the scattering phase shifts from the finite-volume energy levels,
measured in lattice QCD.
The method has been generalized to the case of moving frames,
particles with spin and coupled two-body channels (see here~\cite{Rummukainen:1995vs,Lage:2009zv,Bernard:2010fp,He:2005ey,Liu:2005kr,Hansen:2012tf,Briceno:2012yi,Li:2012bi,Guo:2012hv,Leskovec:2012gb,Kim:2005gf,Gockeler:2012yj} for a representative list of
references). In all cases, the formalism is based on the fundamental assumptions, namely:
\begin{itemize}
\item
  The interactions between particles are short-range. The relation $R/L\ll 1$ holds, where
  $R$ is the characteristic range of interaction and $L$ denotes the size of the cubic box
  (the spatial extension of the lattice) in which the system is placed. The quantity $R$ is
  typically given by the inverse of the lightest mass in the theory, $R\sim M^{-1}$.

\item
  \begin{sloppypar}
  Owing to the condition $R/L\ll 1$, the polarization corrections, proportional to
  $\exp(-L/R)$, are strongly suppressed and can be neglected. This allows one to write
  down an equation (referred to as the L\"uscher equation or the two-body
  quantization condition), which determines the finite-volume
  spectrum in terms of the observables ($S$-matrix elements) only. The details of the
  short-range interactions do not matter.
  \end{sloppypar}

  \item
    Again, owing to the condition $R/L\ll 1$, the partial-wave mixing is small, and it is
    possible to truncate higher partial waves in the L\"uscher equation.

  \end{itemize}

  Obviously, the condition $R/L\ll 1$ is violated, when the scattering in the presence of the
  electromagnetic interactions is considered (QCD+photons on the lattice). Moreover,
  at physical quark masses the pions are rather light, which leads to problems in the
  study of nucleon-nucleon interactions on the lattice. Namely, as explicitly
  demonstrated in the recent paper~\cite{Meng:2021uhz}, the partial-wave mixing
  at the physical point in the L\"uscher equation for the $NN$ scattering is indeed
  substantial. A closely related problem is the appearance of the so-called $t$-channel
  (left-hand) cut in the $NN$ {partial-wave scattering amplitudes},
  running from negative infinity
  till $s=(2m_N)^2-M_\pi^2$ along the real axis in the complex $s$-plane, see, e.g.,
  \cite{Raposo:2023nex,Raposo:2023oru}
 (Here, $m_N$ and $M_\pi$ denote the nucleon and the pion masses, respectively.).  
  Since $M_\pi\ll m_N$ in nature,
  the gap between the $t$-channel cut and the right-hand unitarity cut is very small.
  On the other hand, as seen in the latest studies of
  the $NN$ scattering on the lattice, many energy levels are
  located on the $t$-channel cut (see, e.g., Fig.~3 in Ref.~\cite{Green:2021qol}). The
  standard L\"uscher approach is obviously not applicable in this case. Note also that
  $NN$ scattering is not only the physically interesting process where these problems
  emerge. For example, in the description of the  $T_{cc}(3875)^+$ state the same
  problem shows up in full glory. Moreover, it can be seen that
  the structure of the singularities
  is completely different in the two cases $M_\pi<M_{D^*}-M_D$ and $M_\pi>M_{D^*}-M_D$,
  and hence reveals a
  critical dependence on the values of quark masses used in the lattice calculations~\cite{Du:2023hlu}.

  As a remedy to the above problems, the authors of Ref.~\cite{Meng:2021uhz}
  have advocated solving the quantization condition in the
  three-dimensional plane-wave basis, in order to determine
  the parameters of the effective chiral Lagrangian
  directly from the fit to the lattice energy levels.\footnote{Albeit all calculations in
  Ref.~\cite{Meng:2021uhz} have been carried out in the framework of the chiral EFT,
  one could choose here any EFT that allows a {\em controllable} expansion of
  the scattering amplitude in some small parameter(s) in the energy region of interest.} A similar
  method has recently been applied
  to the analysis of lattice data aimed at the extraction of the $T_{cc}(3875)^+$
  pole~\cite{Du:2023hlu,Meng:2023bmz}.\footnote{Note also that earlier the same method has been used in the continuum, namely, for the study of the energy dependence of the $NN$ scattering amplitude at non-physical quark masses as well~\cite{Baru:2015ira,Baru:2016evv}.}
  This proposal solves the problem
  in principle and translates the output of lattice calculations into the parameters of
  the effective Hamiltonian. The phase shifts and other infinite-volume observables
  are then obtained by solving integral equations in the infinite volume.
  As a side remark, note that a (conceptually) 
  similar solution  is adopted in all formulations of the three-body quantization
condition~\cite{Hansen:2014eka,Hansen:2015zga,Hammer:2017uqm,Hammer:2017kms,Mai:2017bge,Mai:2018djl}, where no other approach to the problem has been found so far.
  However, in the much simpler two-body case, one is tempted to look further for the
  alternatives (similar to the L\"uscher formula)
  which directly express the finite-volume two-body spectrum in terms of the
  physical observables.

  An alternative solution, which has been suggested recently
  \cite{Raposo:2023nex,Raposo:2023oru}, is based on splitting the
  hadron interactions into the long-range and short-range components that are then
  treated separately. In this respect, the approach described in these papers is
  conceptually close to the one pursued in the present work. While a detailed
  comparison of all existing approaches will be given at the end of this work, we
  still mention here the most important difference between
  Refs.~\cite{Raposo:2023nex,Raposo:2023oru} and our paper. 
  Namely, in Refs.~\cite{Raposo:2023nex,Raposo:2023oru} an auxiliary on-shell
  $K$-matrix $\bar K^{\sf os}$ has been introduced, which has to be determined from the
  fit to the lattice data. Once this is done, one should solve the integral equations,
  in order to arrive at the physical amplitudes. So, it is essentially a two-step process.
  In our paper an analog of this auxiliary
  $K$-matrix is introduced as well. However, its relation to the physical amplitude
  has an algebraic form and there is no need to solve integral equations. From this point
  of view, our approach is closer to the original L\"uscher single-step
  formalism than the approach
  described in Refs.~\cite{Raposo:2023nex,Raposo:2023oru}.

Note also that in Ref.~\cite{Hansen:2024ffk}, in the context of the study of the $T_{cc}(3875)^+$ meson,
it was proposed to solve the problem of the $t$-channel cut by writing down three-body
equations, even in case of a stable $D^*$ meson. Below, we shall consider this proposal in more
detail. Here we only note that, in latter case, it is very close to the solution proposed in Refs.~\cite{Meng:2021uhz,Du:2023hlu,Meng:2023bmz}.

  For completeness, we also note that,
  the two-body quantization condition (as well as the two-body Lellouch-L\"uscher formula) in the presence of the Coulomb force
  was considered in Refs.~\cite{Beane:2014qha,Cai:2018why,Christ:2021guf}.
  In the final expressions, the Coulomb potential has been treated perturbatively in the fine structure constant. Such a treatment can be justified for sufficiently small values of the box size $L$.

  The aim of the present paper is to address the problem of the long-range force in the
  L\"uscher equation in a general fashion and to derive a modified L\"uscher equation,
  which has a much larger domain of applicability than the original one.
  To simplify life as much as possible, in this paper we do not consider the
  theories with massless particles -- the inclusion of QED is relegated to future
  publications.
  Furthermore, we ignore purely technical issues like the inclusion of spin, relativistic kinematics or moving
  frames.
  The key observation that allows one to achieve the stated objective
  is that the long-range part of the
  potential, which gives rise to all above problems, is usually well known and can
  be expressed in terms of few parameters that can be accurately measured on the lattice.
    The short-range part of potential is unknown and should be fitted to the lattice data on
  the two-body energy levels by using the modified L\"uscher equation. 

  We shall see below that the method to achieve the above goal is to reformulate the so-called modified
  effective-range expansion (MERE)~\cite{vanHaeringen:1981pb} in a finite volume.
  To this end, in Sect.~\ref{sec:infinite} we invest a certain effort to relate MERE to
  the non-relativistic effective theory (NREFT) framework along the lines described in
  Ref.~\cite{Kong:1999sf} and discuss, in particular, the inclusion of the non-derivative
  couplings which were omitted in Ref.~\cite{Kong:1999sf}. The latter framework can be
  directly recast in a finite volume, as done in Sect.~\ref{sec:finite}, and leads
  to a modified quantization condition with the long-range part split off.
  Section~\ref{sec:comparison} is dedicated to the comparison of our approach to
  alternative ones known in the literature. The numerical implementation of the
  proposed framework constitutes a separate piece of work and will not be considered
  here.

  \section{Modified effective-range expansion in the effective field
    theory framework}
  \label{sec:infinite}

\subsection{Modified effective range expansion}
  
  In Ref.~\cite{vanHaeringen:1981pb}, van Haeringen and Kok consider
  a non-relativistic scattering problem
  for a sum of two local, rotationally invariant potentials:
  \eq
  V(r)=V_L(r)+V_S(r)\, .
  \en
  Here, $V_L(r)$ and $V_S(r)$ denote the long-range and short-range parts of the
  potential, res\-pec\-ti\-ve\-ly. Due to the long-range nature of the full potential,
  the effective-range expansion in the partial-wave with the angular
  momentum $\ell$,
  \eq\label{eq:ERE}
  q^{2\ell+1}\cot\delta_\ell(q)=-\frac{1}{a_\ell}+\frac{1}{2}\,r_\ell q^2+\cdots\, ,
  \en
  has a very small radius of convergence. This happens, e.g., if
  the effective range
  $r_\ell$ and the subsequent coefficients (shape parameters) are unnaturally
  large. However, for a general long-range potential it is also possible
  that the radius of convergence is zero, as in the case of a Coulomb potential.

  Further, the authors define the function
  \eq\label{eq:KlM}
  K_\ell^M(q^2)=M_\ell(q)+\frac{q^{2\ell+1}}{|f_\ell(q)|^2}\,
  \left(\cot(\delta_\ell(q)-\sigma_\ell(q))-i\right)\, .
    \en
    Here, $\delta_\ell(q),\sigma_\ell(q)$ denote, respectively, the full
    phase shift and the phase
    shift in the problem with the long-range potential $V_L(r)$ only
    (i.e., setting $V_S(r)=0$).
    Furthermore, 
    \eq\label{eq:Mell}
    M_\ell(q)=\frac{1}{\ell!}\,\left(-\frac{iq}{2}\right)^\ell
  \lim_{r\to 0}\frac{d^{2\ell+1}}{dr^{2\ell+1}}\,r^\ell\frac{f_\ell(q,r)}{f_\ell(q)}\, ,
  \en
  where $f_\ell(q,r)$ is the Jost solution in the case $V_S(r)=0$, and
\eq
f_\ell(q)=\frac{q^\ell e^{-i\ell\pi/2}(2\ell+1)}{(2\ell+1)!!}\,
\lim_{r\to 0}r^\ell f_\ell(q,r)\, .
\en
  The main result of Ref.~\cite{vanHaeringen:1981pb} consists in
  demonstrating the fact that the quantity $K_\ell^M(q^2)$, defined
  by Eq.~(\ref{eq:KlM}), is a polynomial in the variable $q^2$, with a radius
  of convergence much larger than the original version of the effective-range expansion, displayed in Eq.~(\ref{eq:ERE}).

  The derivation given in Ref.~\cite{vanHaeringen:1981pb}, however, has a
  caveat that has been
  briefly mentioned already in the same paper and was discussed in more detail in
  Ref.~\cite{Badalian:1981xj}. Namely, the quantity $M_\ell(q)$ is well-defined,
  if and only if the potential $V_L(r)$ is regular enough at the origin, so that
  $r^{-2\ell}V_L(r)$ stays analytic at $r=0$. The class of such potentials is
  termed ``superregular'' in Ref.~\cite{Badalian:1981xj}. The usual Coulomb or Yukawa
  potentials do not belong to this class, even for S-wave scattering.

  If one is dealing with potentials which are not superregular, one has to use certain
  convention on top of Eq.~(\ref{eq:Mell}) in order to define the quantity $M_\ell(q)$.
  This is nothing but a renormalization prescription that has to be imposed on the
  ultraviolet-divergent loop containing an arbitrary number of instantaneous ``exchanges''
  corresponding to the potential $V_L(r)$. A non-trivial problem consists in a
  mathematically
  consistent formulation of the renormalization prescription, using the same language as
  used in the derivation of Eqs.~(\ref{eq:KlM}) and (\ref{eq:Mell}). There exists a
  well-known exact solution of the problem for any $\ell$ in case of the Coulomb interaction
  that is given in the textbooks, see, e.g.,~\cite{Goldberger-Watson}. The solution for
  a general potential which is less singular than $r^{-3/2}$
is discussed in Ref.~\cite{Badalian:1981xj} albeit only in the case $\ell=0$ and, roughly
  speaking, boils
  down to a subtraction of the $q$-independent (divergent) constant from $M_\ell(q)$.
  We are not, however, aware
  of the discussion of the case $\ell\neq 0$ in the literature. The term that is divergent at $r\to 0$
  can be identified in this case as well. However, its coefficient is $q$-dependent,
  in general, and the challenge consists in showing that this coefficient is a low-energy
  polynomial in $q^2$ in the wide region determined by the heavy scale.
  For this reason, in the present paper we adopt a different strategy. Namely, we
  truncate the partial-wave expansion $\ell\leq \ell_{\sf max}$ from the beginning
  and regularize the potential $V_L(r)$ in order to render it superregular. For example,
  a kind of the Pauli-Villars regularization will perfectly do the job in case of the Yukawa
  potential we shall be primarily dealing with:
  \eq
  V_L(r)=\frac{ge^{-M_\pi r}}{r}\to \frac{ge^{-M_\pi r}}{r}-\sum_{i=1}^{2\ell_{\sf max}+1}
  c_i\frac{ge^{-M_i r}}{r}\, .
  \en
  Here, $M_i=n_iM$, where $M$ denotes a typical heavy scale of the
  theory (determined, for instance, by the inverse range in $V_S(r)$), whereas $n_i$
  are numbers of order unity. The requirement that the first $2\ell_{\sf max}+1$ terms in the Laurent expansion vanish leads to the following linear system of equations:
  \eq
  1&=&\sum_{i=1}^{2\ell_{\sf max}+1}c_i\, ,
  \nonumber\\[2mm]
  \frac{M_\pi}{M}&=&\sum_{i=1}^{2\ell_{\sf max}+1}c_in_i\, ,
  \nonumber\\[2mm]
  &&\cdots\, ,
  \nonumber\\[2mm]
  \left(\frac{M_\pi}{M}\right)^{2\ell_{\sf max}}&=&
  \sum_{i=1}^{2\ell_{\sf max}+1}c_in_i^{2\ell_{\sf max}}\, .
  \label{eq:superreg}
  \en
  It is straightforwardly seen that the regularized potential is indeed superregular for
  all $\ell\leq\ell_{\sf max}$. Note, however, that the $2\ell_{\sf max}+1$ equations in Eq.~(\ref{eq:superreg}) do not determine the  $n_i$ and $c_i$ uniquely.
  The constants $n_i$ have to be picked such that the masses $M_i=n_iM$ are of order $M$ and different from each other. This ensures that
  Eq.~(\ref{eq:superreg}) has a solution and no unnaturally large coefficients
  $c_i$ emerge.

  Finally, in the splitting $V(r)=V_L(r)+V_S(r)$, one could modify both $V_L(r)$ and $V_S(r)$, adding and subtracting the same string of the short-range Yukawa terms. This will render
  $ V_L(r)$ superregular and will not change the interpretation of $V_S(r)$ as a short-range
  potential. {Hence, the above regularization
 serves solely the purpose of recasting the potential into
    the form that obeys the requirements of Ref.~\cite{vanHaeringen:1981pb}.}

  \subsection{NREFT framework}
  In the literature, there have been several attempts to reformulate the modified
  effective-range expansion in the effective field theory language~\cite{Kong:1999sf,Beane:2014qha,Birse:1998dk,Steele:1998zc}. We shall mainly follow
  the path outlined in these papers and derive an analog of
  Eq.~(\ref{eq:KlM}) in the effective field theory setting. In order to do
  this, we recall that, in the non-relativistic effective field theory, the scattering amplitude is merely a solution of the Lippmann-Schwinger (LS) equation with the potential determined by a matrix element of the interaction Lagrangian between the free two-particle states. We still assume that the potential is a sum of long-range and short-range parts, but do not assume anymore that the potential
  is local. The short-range potential in momentum space is a familiar
  low-energy polynomial. Its partial-wave expansion can be written in the following
  form
  \eq\label{eq:VS}
\langle {\bf p}|V_S|{\bf q}\rangle=4\pi\sum_{\ell m}Y_{\ell m}(\hat p)
V_S^\ell(p,q)Y_{\ell m}^*(\hat q)\, .
\en
Here,
\eq\label{eq:VSell}
V_S^\ell=(pq)^\ell\sum_{a=0}^\infty \sum_{b=0}^aC_\ell^{ab}(p^2)^b(q^2)^{a-b}\, ,
\quad\quad C_\ell^{ab}=C_\ell^{ba}\, .
\en
Furthermore, $p=|{\bf p}|$, $\hat p$ denotes a unit vector in direction of
    ${\bf p}$, and $Y_{\ell m}(\hat p)$ are the spherical harmonics.
Writing down explicitly the first few terms in the potential, one gets
\eq\label{eq:polynomial}
  \langle {\bf p}|V_S|{\bf q}\rangle=C_0^{00}+3C_1^{00}{\bf p}{\bf q}+C_0^{10}({\bf p}^2+{\bf q}^2)+\cdots\, ,
  \en
  which in the position space corresponds to a sum of a $\delta$-like potential
  and the derivatives thereof. The long-range potential might be taken to be the regularized Yukawa
  potential, corresponding to an exchange of a light particle, see above. In any case, it is assumed
  to be local.
  Furthermore, ultraviolet divergences will be present in the
  LS equation, in general. We assume that these divergences are regularized
  and renormalized in a standard fashion (say, the power-divergence subtraction
  (PDS) scheme, bearing the case of $NN$ scattering in mind).
  Since the presence of a long-range force non-trivially affects
  only the infrared behavior of the theory, it is expected that the issue of
  renormalization is inessential in the present context. To simplify things, one could also merely assume that the momentum cutoff is performed at a very large value
  $\Lambda$, and the $\Lambda$-dependent effective couplings are adjusted
  order by order to
  reproduce the behavior of the $S$-matrix elements at low momenta.

\begin{figure}[tbp]
\centerline{
    \includegraphics[width=.8\textwidth]{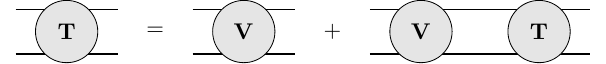}}
    \caption{Lippmann-Schwinger equation for the full $T$ matrix (circle marked $T$)
    in momentum space. The circle marked $V$ indicates the full potential while the pair of internal solid lines represent the free two-particle Green function $G_0$. Closed loops imply a momentum integration. }
    \label{fig:LSeq}
\end{figure}
The fully off-shell LS equation for the $T$ matrix 
is illustrated in Fig.~\ref{fig:LSeq}. 
  The corresponding 
  integral equation in momentum space is given by
  \eq
  T({\bf p},{\bf q};q_0^2+i\varepsilon)=V({\bf p},{\bf q})
  +\int\frac{d^3{\bf k}}{(2\pi)^3}\,
  \frac{V({\bf p},{\bf k})T({\bf k},{\bf q};q_0^2+i\varepsilon)}{{\bf k}^2-q_0^2 -i\varepsilon}\, ,
  \en
  where the explicit expression for the two-particle Green function 
  \eq
    \langle {\bf p}|G_0(q_0^2+i\varepsilon)|{\bf q}\rangle
    =\frac{(2\pi)^3\delta^3({\bf p}-{\bf q})}{{\bf p}^2-q_0^2-i\varepsilon}\, .
    \en
  was used and
  the regularization of the momentum integration with the cutoff $\Lambda$ is left implicit.
  The partial-wave amplitudes are defined as follows:
  \eq
  T({\bf p},{\bf q};q_0^2+i\varepsilon)=4\pi\sum_{\ell m}
  Y_{\ell m}(\hat p) T_\ell(p,q;q_0^2+i\varepsilon)Y_{\ell m}^*(\hat q)\, .
    \en
   The phase shift is related to the on-shell partial-wave amplitudes via:
    \eq\label{eq:T_ell}
    T_\ell(q_0,q_0;q_0^2+i\varepsilon)\doteq T_\ell(q_0)=\frac{4\pi}{q_0\cot\delta_\ell(q_0)-iq_0}\, .
    \en
    
\begin{figure}[tbp]
\centerline{
    \includegraphics[width=.8\textwidth]{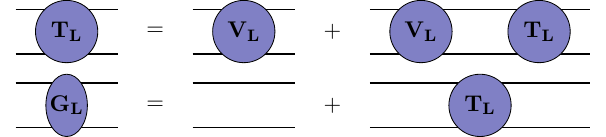}}
    \caption{Definition of the scattering amplitude $T_L$ (first line) and Green function $G_L$ (second line) for the long-range interaction only. Further notation as in Fig.~\ref{fig:LSeq}}
    \label{fig:longrange}
\end{figure}

\begin{figure}[tbp]
\centerline{
    \includegraphics[width=.8\textwidth]{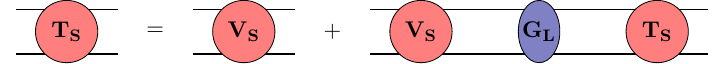}}
    \caption{Diagrams for the short-range $T$ matrix $T_L$. Note that the loop integration now involves $G_L$ instead of $G_0$. Further notation as in Fig.~\ref{fig:LSeq}}
    \label{fig:TL}
\end{figure}

\begin{figure}[tbp]
\centerline{
    \includegraphics[width=.8\textwidth]{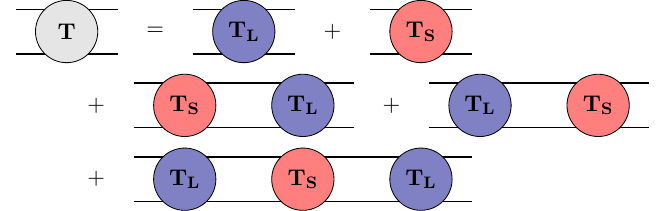}}
    \caption{Diagrams for the full $T$ matrix expressed through $T_L$ and $T_S$. Further notation as in Fig.~\ref{fig:LSeq}}
    \label{fig:fullT}
\end{figure}

    Next, we split the full potential into the long- and short range parts,
    $V=V_L+V_S$, and define the scattering amplitude $T_L$ and the Green function
    $G_L$ for the long-range potential only.
    This construction is shown diagrammatically in Fig.~\ref{fig:longrange}.
The first line gives the LS equation for $T_L$, while the second line gives the expression for $G_L$ in terms of $G_0$ and $T_L$.  Both quantities are needed for the NREFT formulation of the modified effective range expansion. The explicit expressions read   
    \eq\label{eq:longrangeonly}
    T_L(q_0^2+i\varepsilon)&=&V_L+V_LG_0(q_0^2+i\varepsilon)T_L(q_0^2+i\varepsilon)\, ,
    \nonumber\\[2mm]
    G_L(q_0^2+i\varepsilon)&=&G_0(q_0^2+i\varepsilon)+G_0(q_0^2+i\varepsilon)
    T_L(q_0^2+i\varepsilon)G_0(q_0^2+i\varepsilon)\, .
    \en
Moreover, we define the short-range $T$ matrix
$T_S$, 
\eq\label{eq:T_S}
    T_S(q_0^2+i\varepsilon)=V_S+V_SG_L(q_0^2+i\varepsilon)T_S(q_0^2+i\varepsilon)\, .
    \en
which is shown diagrammatically in 
Fig.~\ref{fig:TL}.  Note that the loop integration now involves $G_L$ instead of $G_0$.
Using these definitions, the full $T$ matrix can be expressed as
    \eq\label{eq:twopotential}
    &&\hspace*{-1.5cm}T(q_0^2+i\varepsilon)\,=\,T_L(q_0^2+i\varepsilon)
\nonumber\\[2mm]
    &+& (1+T_L(q_0^2+i\varepsilon)G_0(q_0^2+i\varepsilon))
    T_S(q_0^2+i\varepsilon)(G_0(q_0^2+i\varepsilon)T_L(q_0^2+i\varepsilon)+1)\, ,
    \en
    which is illustrated in Fig.~\ref{fig:fullT}.
This amounts to adding $T_L$ and $T_S$ dressed by $T_L$ in all possible ways.

    The above expressions are of course familiar from the theory of scattering
    on two potentials. Note that we have used operator notation
    in Eqs.~(\ref{eq:longrangeonly}, \ref{eq:T_S}, \ref{eq:twopotential}) to
    keep the notation clear. The
    integrations over intermediate states will only be shown explicitly in the
    following when required for clarity.

In order to simplify life further, we assume that the long-range potential
is repulsive and does not create bound states. Then, the spectral
representation of the long-range Green function takes the form:
\eq
 \langle {\bf p}|G_L(q_0^2+i\varepsilon)|{\bf q}\rangle
 =\int\frac{d^3{\bf k}}{(2\pi)^3}\,
 \frac{\langle {\bf p}|\psi^{(\pm)}_{\bf k}\rangle
   \langle \psi^{(\pm)}_{\bf k}|{\bf q}\rangle}
      {{\bf k}^2-q_0^2-i\varepsilon}\, ,
      \en
      where $\psi^{(\pm)}_{\bf k}$ denote the eigenfunctions of the Hamiltonian
      $H_L=H_0+V_L$, corresponding to the eigenvalue ${\bf k}^2$, and $(\pm)$
      specifies outgoing/ingoing boundary conditions on the wave function.
      These wave functions can be constructed with the use of the
      M\o{}ller operators:
\eq\label{eq:onshellpsi}
|\psi^{(\pm)}_{\bf k}\rangle&=&(1+G_0(k^2\pm i\varepsilon)T_L(k^2\pm i\varepsilon))|{\bf k}\rangle\doteq \Omega(k^2\pm i\varepsilon)|{\bf k}\rangle\, ,
\nonumber\\[2mm]
\langle\psi^{(\pm)}_{\bf k}|&=&\langle {\bf k}|(1+T_L(k^2\mp i\varepsilon)G_0(k^2\mp i\varepsilon))\doteq \langle{\bf k}|\Omega^\dagger(k^2\pm i\varepsilon)\, .
\en
Now, let us consider the Born series for the quantity
\eq
&&\langle{\bf p}|\Omega^\dagger(q_0^2-i\varepsilon)T_S(q_0^2+i\varepsilon)\Omega(q_0^2+i\varepsilon)|{\bf q}\rangle
\nonumber\\[2mm]
&=&\int_{{\bf k}_1, {\bf k}_2}
\langle{\bf p}|\Omega^\dagger(q_0^2-i\varepsilon)|{\bf k}_1\rangle
\langle{\bf k}_1|V_S|{\bf k}_2\rangle
\langle{\bf k}_2|\Omega(q_0^2+i\varepsilon)|{\bf q}\rangle
\nonumber\\[2mm]
&+&\int_{{\bf k}_1, {\bf k}_2, {\bf k}_3, {\bf k}_4,{\bf l}}
\langle{\bf p}|\Omega^\dagger(q_0^2-i\varepsilon)|{\bf k}_1\rangle
\langle{\bf k}_1|V_S|{\bf k}_2\rangle\langle{\bf k}_2|\psi^{(+)}_{\bf l}\rangle
\frac{1}{{\bf l}^2-q_0^2-i\varepsilon}\,
\langle\psi^{(+)}_{\bf l}|{\bf k}_3\rangle
\nonumber\\[2mm]
&&\hspace*{.5cm}\times\,\langle{\bf k}_3|V_S|{\bf k}_4\rangle
\langle{\bf k}_4|\Omega(q_0^2+i\varepsilon)|{\bf q}\rangle+\cdots\, ,
\en
where we have used the abbreviation
\eq
\int_{{\bf k}_1}\equiv \frac{d^3{\bf k}_1}{(2\pi)^3}\,.
\en
On the energy shell $p^2=q^2=q_0^2$, the above expression simplifies to
\eq\label{eq:B}
&&\langle{\bf p}|\Omega^\dagger(q_0^2-i\varepsilon)T_S(q_0^2+i\varepsilon)\Omega(q_0^2+i\varepsilon)|{\bf q}\rangle
\nonumber\\[2mm]
&=&\int_{{\bf k}_1, {\bf k}_2}
\langle\psi^{(-)}_{\bf p}|{\bf k}_1\rangle
\langle{\bf k}_1|V_S|{\bf k}_2\rangle\langle {\bf k}_2|\psi^{(+)}_{\bf q}\rangle
\nonumber\\[2mm]
&+&\int_{{\bf k}_1, {\bf k}_2, {\bf k}_3, {\bf k}_4,{\bf l}}
\langle\psi^{(-)}_{\bf p}|{\bf k}_1\rangle
\langle{\bf k}_1|V_S|{\bf k}_2\rangle\langle{\bf k}_2|\psi^{(+)}_{\bf l}\rangle
\frac{1}{{\bf l}^2-q_0^2-i\varepsilon}\,
\langle\psi^{(+)}_{\bf l}|{\bf k}_3\rangle
\nonumber\\[2mm]
&&\hspace*{.5cm}\times\,\langle{\bf k}_3|V_S|{\bf k}_4\rangle
\langle{\bf k}_4|\psi^{(+)}_{\bf q}\rangle+\cdots\, .
\nonumber\\
\en
The partial-wave expansion of the asymptotic wave functions is defined as follows:
\eq
\langle{\bf p}|\psi^{(\pm)}_{\bf k}\rangle=
\sum_{\ell m}Y_{\ell m}(\hat p)\psi^{(\pm)}_\ell(k,p)Y_{\ell m}^*(\hat k)\, .
  \en
  Furthermore,
  \eq
  \left(\psi^{(-)}_\ell(k,p)\right)^*
  =e^{2i\sigma_\ell(k)}\left(\psi^{(+)}_\ell(k,p)\right)^*\, ,
  \en
  where $\sigma_\ell(k)$ denotes the scattering phase shift in case of
  the long-range potential only. Now, the partial-wave expansion
  of the quantity defined in Eq.~(\ref{eq:B}) is given by
  \eq\label{eq:expB}
\langle{\bf p}|\Omega^\dagger(q_0^2-i\varepsilon)T_S(q_0^2+i\varepsilon)\Omega(q_0^2+i\varepsilon)|{\bf q}\rangle 
=4\pi\sum_{\ell m}Y_{\ell m}(\hat p)e^{2i\sigma_\ell(q_0)}B_\ell(q_0)Y_{\ell m}^*(\hat q)\, ,
  \en
  where
    \eq\label{eq:Beq}
  B({\bf p},{\bf q};q_0^2+i\varepsilon)=\tilde V_S({\bf p},{\bf q})
  +\int\frac{d^3{\bf k}}{(2\pi)^3}\,
  \frac{\tilde V_S({\bf p},{\bf k})B({\bf k},{\bf q};q_0^2+i\varepsilon)}{{\bf k}^2-q_0^2-i\varepsilon}\, ,
  \en
  \eq\label{eq:tildeV}
  \tilde V_S({\bf p},{\bf q})
  =\langle\psi^{(+)}_{\bf p}|V_S|\psi^{(+)}_{\bf q}\rangle\, ,
  \en
  and $B_\ell(q_0)$ is equal to the partial-wave amplitude
  $B_\ell(p,q;q_0^2)$ on the energy shell $p^2=q^2=q_0^2$.
  In analogy to Eq.~(\ref{eq:T_ell}), one may write
    \eq\label{eq:B_ell}
    B_\ell(q_0)=\frac{4\pi}{q_0\cot\tilde\delta_\ell(q_0)-iq_0}\, .
    \en  
    Using Eqs.~(\ref{eq:twopotential}) and (\ref{eq:expB}), one finally gets:
    \eq
    \tilde\delta_\ell(q_0)=\delta_\ell(q_0)-\sigma_\ell(q_0)\, .
    \en

    \subsection{Non-derivative interactions}

    Let us first
    restrict ourselves to $\ell=0$ and assume that only the coupling
    $C_0^{00}$ is different from zero. Then, the potential $\tilde V_S$ is
    separable:
    \eq\label{eq:tildeVS}
    \tilde V_S({\bf p},{\bf q})=\left(\tilde\psi^{(+)}_{\bf p}(0)\right)^*
    C_0^{00} \tilde\psi^{(+)}_{\bf q}(0)\, .
    \en
    Here, $\tilde\psi^{(+)}_{\bf q}({\bf r})$ stands for the wave function in
    the coordinate space. Then, on the energy shell $|{\bf q}|=q_0$, the S-wave amplitude
    takes the form
    \eq\label{eq:B0}
    B_0(q_0)=\frac{|\tilde\psi^{(+)}_{\bf q}(0)|^2}{(C_0^{00})^{-1}-\langle G_L^0(q_0)\rangle}\, ,
    \en
    where
    \eq
    \langle G_L^0(q_0)\rangle=\int\frac{d^3{\bf k}}{(2\pi)^3}\,\frac{|\tilde\psi^{(+)}_{\bf k}(0)|^2}{{\bf k}^2-q_0^2-i\varepsilon}\, .
    \en
    Let us now assume that the conditions of Ref.~\cite{vanHaeringen:1981pb}
    are fulfilled and, namely, the long-range potential $V_L$ is local.
    It is convenient to define the partial-wave expansion
    of the Green function in the momentum/coordinate spaces
    as follows:
    \eq
    \langle {\bf p}|G_L(q_0^2+i\varepsilon)|{\bf q}\rangle
    &=&4\pi\sum_{\ell m}\mathscr{Y}_{\ell m}({\bf p})G_L^\ell(p,q;q_0^2+i\varepsilon)
    \mathscr{Y}_{\ell m}^*({\bf q})\, ,
\nonumber\\[2mm]
    \langle {\bf r}|G_L(q_0^2+i\varepsilon)|{\bf w}\rangle
    &=&4\pi\sum_{\ell m}\mathscr{Y}_{\ell m}({\bf r})\tilde G_L^\ell(r,w;q_0^2+i\varepsilon)
    \mathscr{Y}_{\ell m}^*({\bf w})\, ,
    \en
    with $\mathscr{Y}_{\ell m}({\bf z})=z^\ell Y_{\ell m}(\hat z)$. Then,
\eq
    \langle G_L^\ell(q_0)\rangle=\lim_{r,w\to 0}\tilde G_L^\ell(r,w;q_0^2+i\varepsilon)\, .
    \en
    Using the result of Appendix
    \ref{sec:calcGreen}, we can write
\eq\label{eq:GLell}
\langle G_L^\ell(q_0)\rangle=\frac{1}{4\pi\left((2\ell+1)!!\right)^2}\,M_\ell(q_0)+\mbox{real polynomial in $q_0^2$}\, ,
  \en
  where $M_\ell(q_0)$ is given by Eq.~(\ref{eq:Mell}). The real polynomial can be safely dropped as it corresponds to
  the choice of the renormalization prescription.
  
  {Next, we express the wave function at the origin, which appears in Eq.~(\ref{eq:B0}),
  through the scattering wave function~$\phi_0(k,r)\doteq \phi_0(k,k,r)$
  defined in Eq.~(\ref{eq:A3}).} Using
  \eq
  |\tilde\psi_{\bf k}(0)|=\lim_{r\to 0}\left|\frac{\phi_0(k,r)}{kr}\right|=\frac{1}{|f_0(k)|}
  \en
along with Eqs.~(\ref{eq:B_ell}) and (\ref{eq:B0}), we obtain
\eq
  \frac{4\pi}{C_0^{00}}=M_0+\frac{q_0}{|f_0(q_0)|^2}\,(\cot\tilde\delta_0(q_0)-i)\,
  \en  
for the S-wave phase shift.
 The essence of the modified effective-range expansion is now crystal clear: it has a larger radius of convergence which is governed by  the short-range potential only.

\subsection{Derivative interactions}
\label{sec:derivative}

Consider now the situation when the matrix element of the potential $V_S$ is a generic
low-energy polynomial defined in Eq.~(\ref{eq:polynomial}). This is no more true
for the potential $\tilde V_S({\bf p},{\bf q})$, defined in Eq.~(\ref{eq:tildeV}). Here,
we wish to address the structure of the latter in more detail. The partial-wave expansion
of the $V_S$ is given in Eq.~(\ref{eq:VS}).
Convoluting this equation with the wave functions,
integrals of the following type emerge
\eq\label{eq:Aelln}
A_\ell^a&=&\int\frac{d^3{\bf p}}{(2\pi)^3}\,
\mathscr{Y}_{\ell m}^*({\bf p})({\bf p}^2)^a\langle{\bf p}|\psi^{(+)}_{\bf k}\rangle\, .
\en
One can now use the identity
\eq({\bf p}^2)^a=({\bf p}^2-{\bf k}^2+{\bf k}^2)^a
=({\bf p}^2-{\bf k}^2)^a+a({\bf p}^2-{\bf k}^2)^{a-1}{\bf k}^2+\cdots\, ,
\en
and rewrite Eq.~(\ref{eq:Aelln}) as
\eq
A_\ell^a=\lim_{{\bf r}\to 0}
\mathscr{Y}_{\ell m}^*(i\nabla)\sum_{b=0}^a\frac{(-1)^{a-b} a!}{b!(a-b)!}\,
({\bf k}^2)^b({\bf k}^2+\triangle)^{a-b}\tilde\psi^{(+)}_{\bf k}({\bf r})\, .
\en
Here, as in Ref.~\cite{vanHaeringen:1981pb}, it is assumed that the long-range potential
$V_L(r)$ is local and spherically symmetric. Furthermore, consider the case $\ell=0$ first.
Using the Schr\"odinger equation, one then gets
\eq
({\bf k}^2+\triangle)\psi^{(+)}_{\bf k}({\bf r})=\mbox{const}\cdot V_L(r)\psi^{(+)}_{\bf k}({\bf r})\to\mbox{const}\cdot V_L(0)\psi^{(+)}_{\bf k}(0)\, ,\quad\mbox{as}~{\bf r}\to 0\, .
\en
In the case of a regularized Yukawa coupling, the quantity $V_L(0)$ is finite.

Acting now with the operator $({\bf k}^2+\triangle)$ on both sides
of this equation once more,
one gets a term, containing $V_L^2$, as well as terms with the space derivatives acting
on $V_L(r)$. Continuing this operation, we get a string of terms, containing
$\nabla_{i_1}\ldots\nabla_{i_k}V_L(r)$.
Furthermore, owing to the rotational symmetry,
\eq\label{eq:many_nablas}
\lim_{r\to 0}\nabla_{i_1}\ldots\nabla_{i_k}V_L(r)=
\left\{
\begin{array}{ll}
 (\delta_{i_1i_2}\cdots\delta_{i_{k-1}i_k}+\mbox{perm})V_k\, ,
  & \mbox{even }k\, ,\cr
  0\, &\mbox{odd }k\,.
\end{array}
\right.
\en
One could
 stick to the dimensional regularization here, in which all $V_k$ are finite.
Furthermore, in the dimensional regularization, $V_k\sim \mu^k$, where $\mu$ denotes the
small mass scale of the long-distance potential (the pion mass $M_\pi$, in case of
Yukawa interactions).
We remind the reader that we are dealing here with the long-distance (infrared)
problems, for which the details of the ultraviolet renormalization should not matter.

The Kronecker $\delta$-symbols, which are present in
Eq.~(\ref{eq:many_nablas}), can be further contracted with $\nabla_{i_1}\cdots$ acting
on the wave function, turning them into the Laplacians $\triangle$ that can be again
eliminated with the use of the Schr\"odinger equation. At the end of the day, for $\ell=0$,
\eq
A_0^a=
\sum_{b=0}^a
({\bf k}^2)^bh_0^{a-b}\tilde\psi^{(+)}_{\bf k}({\bf 0})\, ,
\en
where the coefficients $h_0^{a-b}$ are expressed through $V_L(0)$ and the derivatives
of the potential at the origin. It is important to mention that the mass scale in the
derivatives is set by the long-range potential and, therefore, the expansion in the
derivatives is converging fast.

Next, consider the case $\ell\neq 0$ and restore the factor
$\mathscr{Y}_{\ell m}^*(i\nabla)$ in the expression for $A_\ell^a$.
This factor contains exactly $\ell$ derivatives which should be commuted through all
potentials to the right. Performing the limit ${\bf r}\to 0$, it is
straightforward to ensure that
\eq
A_\ell^a&=&
\sum_{b=0}^a
({\bf k}^2)^bh_\ell^{a-b}
\lim_{{\bf r}\to 0}\mathscr{Y}_{\ell m}^*(i\nabla)\tilde\psi^{(+)}_{\bf k}({\bf r})
\nonumber\\[2mm]
&\doteq&
4\pi\sum_{b=0}^a
c_\ell ({\bf k}^2)^{a-b}h_\ell^b
\lim_{r\to 0}i^\ell\frac{\phi_\ell(k,r)}{kr^{\ell+1}}\,Y_{\ell m}^*(\hat k)\, ,
\en
where
\eq
c_\ell\delta_{mm'}=\lim_{{\bf r}\to 0}\mathscr{Y}_{\ell m}^*(i\nabla)
\mathscr{Y}_{\ell m'}({\bf r})=\frac{i^\ell \ell! (2\ell+1)}{4\pi}\delta_{mm'}\, .
\en
Furthermore, using
\eq\label{eq:Newton}
\lim_{r\to 0}\frac{\phi_\ell(k,r)}{kr^{\ell+1}}=\frac{k^\ell}{f_\ell(k)(2\ell+1)!!}\, ,
\en
one obtains
\eq\label{eq:tildeVSell}
\tilde V_S^\ell(p,q)&=&\frac{1}{[(2\ell+1)!!]^2f_\ell^*(p)f_\ell(q)}\,(pq)^\ell
\bar V_S^\ell(p,q)\, ,
\nonumber\\[2mm]
\bar V_S^\ell(p,q)&=&\sum_{a=0}^\infty \sum_{b=0}^a\tilde C_\ell^{ab}(p^2)^b(q^2)^{a-b}\, .
\en
In the above expression, the couplings $\tilde C_\ell^{ab}$ are expressed through
$C_\ell^{ab}$ in form of the series in the small scale $\mu$. In other words, no unnaturally large
couplings emerge. This property is crucial for arguing that the sum, given in the above
equation, still represents a low-energy polynomial. To summarize, $\tilde V_S^\ell(p,q)$
unlike $V_S^\ell(p,q)$, is {\em not} a low-energy polynomial. The difference is however
minimal and boils down to the Jost functions that enter the expression as a multiplicative
factor.

At the next step, we carry out the partial-wave expansion in Eq.~(\ref{eq:Beq})
and use the following ansatz for the partial-wave amplitude:
\eq
B_\ell(p,q;q_0^2+i\varepsilon)&=&\frac{1}{[(2\ell+1)!!]^2f_\ell^*(p)f_\ell(q)}\,(pq)^\ell
\bar B_\ell(p,q;q_0^2+i\varepsilon)\, .
\en
This gives
\eq
\bar B_\ell(p,q;q_0^2+i\varepsilon)
=\bar V_S^\ell(p,q)+\int\frac{k^2dk}{2\pi^2}\,\frac{k^{2\ell}}{[(2\ell+1)!!]^2
  |f_\ell(k)|^2}\,\frac{\bar V_S^\ell(p,k)\bar B_\ell(k,q;q_0^2+i\varepsilon)}{k^2-q_0^2-i\varepsilon}\, .
\nonumber\\
\en
Let us now define a new amplitude that obeys an integral equation with a regular
kernel:
\eq
&&\quad\quad R_\ell(p,q;q_0^2)
=\bar V_S^\ell(p,q)
\nonumber\\[2mm]
&+&\int\frac{k^2dk}{2\pi^2}\,\frac{k^{2\ell}}{[(2\ell+1)!!]^2
  |f_\ell(k)|^2}\,\frac{\bar V_S^\ell(p,k) R_\ell(k,q;q_0^2)
  -\bar V_S^\ell(p,q_0) R_\ell(q_0,q;q_0^2)}{k^2-q_0^2}\, .
\en
These two amplitudes on the energy shell are related by
\eq
\bar B_\ell(q_0,q_0;q_0^2+i\varepsilon)&=&R_\ell(q_0,q_0;q_0^2)
+\bar B_\ell(q_0,q_0;q_0^2+i\varepsilon)R_\ell(q_0,q_0;q_0^2)
\nonumber\\[2mm]
&\times&\int\frac{k^2dk}{2\pi^2}\,\frac{k^{2\ell}}{[(2\ell+1)!!]^2
  |f_\ell(k)|^2}\,\frac{1}{k^2-q_0^2-i\varepsilon}\, .
\en
Note that $R_\ell(p,q;q_0^2)$, like $\bar V_S^\ell(p,q)$, is a low-energy polynomial.
Identifying  $K_\ell^M(q_0^2)=\left[R_\ell(q_0,q_0;q_0^2)\right]^{-1}$, we obtain:
\eq
B_\ell(q_0)=\frac{q_0^{2\ell}}{|f_\ell(q_0)|^2[(2\ell+1)!!]^2}\,
\biggl\{
K^M_\ell(q_0,q_0)-\langle G_L^\ell(q_0)\rangle\biggr\}^{-1}\, .
\label{eq:BKG}
\en
Finally, using Eqs.~(\ref{eq:B_ell}) and~(\ref{eq:GLell}), one arrives at the modified
effective-range expansion as given in Eq.~(\ref{eq:KlM}), with $K_\ell^M(q_0^2)$
being a low-energy polynomial.

To summarize, using effective field theory methods, we have rederived the modified
effective range expansion formula of Ref.~\cite{vanHaeringen:1981pb}, where the effects
of the long-range interactions are separated and included in the functions $f_\ell(q)$
and $M_\ell(q)$ that do not depend on the short-range potential $V_S$. This neat separation
is, however, based on the assumption that the long-range potential $V_L(r)$ is local. The
most important cases of the long-range force: the one-pion exchange as well as Coulomb
interactions are exactly of this type. It can be further expected that,
with some effort, the method
could be generalized to the case of a finite sum
$V_1(r)+(\triangle V_2(r)+V_2(r)\triangle)+\nabla V_3(r)\nabla+\ldots$, albeit the final formula probably takes a more complicated form
(Here, the couplings in front of $V_1(r),V_2(r),V_3(r),\ldots$ are assumed to be of natural size.).
In this paper, we are not pursuing this idea further.
On the other hand, a generic non-local long-range potential (say, a separable
potential with a very smooth cutoff) is most likely not amenable to this
kind of treatment at all. In other words, in general, there are two mass scales present
in the potential $V_L$ -- the one associated with the momentum transfer and the
one associated with the
relative momentum in the CM frame, respectively. A long-range potential, in which the former scale
is small whereas the latter scale is of a natural size, can be treated with the method in a similar way as described above.

\section{Modified L\"uscher equation}
\label{sec:finite}

\subsection{Derivation of the quantization condition}

In a finite box,
the Green function $G_L(q_0^2)$, which enters in the equation for $T_S(q_0^2)$, can
be expanded in a sum over all eigenvectors of the Hamiltonian $H_L$
in a finite volume:
\eq\label{eq:spectral}
\langle{\bf p}|G_L(q_0^2)|{\bf q}\rangle
=\sum_n\frac{\langle{\bf p}|\psi_n\rangle\langle\psi_n|{\bf q}\rangle}{q_n^2-q_0^2}\, .
\en
Furthermore, the finite-volume spectrum of the system is determined by the pole
positions of the full $T$-matrix that can be written down in a form of a finite-volume analog of Eq.~(\ref{eq:twopotential}). Using the fact that the poles of $T_L$ will
eventually cancel~\cite{Doring:2009bi} (see also Appendix~\ref{app:cancellation}),
it is straightforward to conclude that the spectrum will be
determined by the poles of $T_S$. Moreover, it is easily seen that no spurious poles
emerge, since the poles that emerge in $G_L$ are shifted by the short-range interaction. 
Using now the basis of eigenfunctions of the Hamiltonian $H_L$ and defining the
quantity
\eq
B^{nm}(q_0^2)=\langle\psi_n|T_S(q_0^2)|\psi_m\rangle\, ,
\en
we get
\eq
B^{nm}=\tilde V_S^{nm}+\sum_k\tilde V_S^{nk}\frac{1}{q_k^2-q_0^2}B^{km}\, ,
\quad\quad 
\tilde V_S^{nm}=\langle\psi_n|V_S|\psi_m\rangle\, .
\en
(The dependence of $B^{nm}$ on $q_0^2$ is suppressed hereafter).
Furthermore, using the partial-wave expansion
\eq
v_{\ell m}^n=\frac{1}{L^3}\,\sum_ {\bf p}\mathscr{Y}_{\ell m}^*({\bf p})
\langle {\bf p}|\psi_n\rangle\, ,
\en
we get
\eq
\langle\psi_n|V_S|\psi_m\rangle=4\pi\sum_{\ell m,\ell'm'}\left(v_{\ell m}^n\right)^*
\bar V_S^\ell(q_n,q_m)\delta_{\ell\ell'}\delta_{mm'}v_{\ell'm'}^m\, .
\en
The next steps in the derivation repeat those in the infinite volume. We use
the following ansatz for the matrix $B$
\eq
B^{nm}=4\pi\sum_{\ell m}\left(v_{\ell m}^n\right)^*
\bar B_{\ell m,\ell'm'}(q_n,q_m;q_0^2)v_{\ell'm'}^m\, ,
\en
and get
\eq
&&\bar B_{\ell m,\ell'm'}(q_n,q_m;q_0^2)=\bar V_S^\ell(q_n,q_m)\delta_{\ell\ell'}\delta_{mm'}
\nonumber\\[2mm]
&+&\sum_k\sum_{\ell''m''}
\bar V_S^\ell(q_n,q_k)\frac{4\pi v_{\ell m}^k\left(v_{\ell''m''}^k\right)^*}{q_k^2-q_0^2}\,\bar B_{\ell''m'',\ell'm'}(q_k,q_m;q_0^2)\, .
\en
Define again
\eq\label{eq:discarded}
&&R_{\ell m,\ell'm'}(q_n,q_m;q_0^2)=\bar V_S^\ell(q_n,q_m)\delta_{\ell\ell'}\delta_{mm'}
+\sum_k\sum_{\ell''m''}
\frac{4\pi v_{\ell m}^k\left(v_{\ell''m''}^k\right)^*}{q_k^2-q_0^2}\,
\nonumber\\[2mm]
&\times&\biggl(\bar V_S^\ell(q_n,q_k)R_{\ell''m'',\ell'm'}(q_k,q_m;q_0^2)
-\bar V_S^\ell(q_n,q_0)R_{\ell''m'',\ell'm'}(q_0,q_m;q_0^2)\biggr)\, .
\en
The infinite-volume limit in this (subtracted) equation can be performed,
and the quantity $R_{\ell m,\ell'm'}(p,q;q_0^2)$ tends to
$\delta_{\ell\ell'}\delta_{mm'}R_\ell(p,q;q_0^2)$ in this limit. On the mass shell,
with $q_n^2=q_m^2=q_0^2$ we, therefore, obtain
\eq
 && \bar B_{\ell m,\ell'm'}(q_0,q_0;q_0^2)=
  \delta_{\ell\ell'}\delta_{mm'}R_\ell(q_0,q_0;q_0^2)
\nonumber\\[2mm]
  &+&\sum_{\ell''m''}  R_\ell(q_0,q_0;q_0^2)
  H_{\ell m,\ell''m''}(q_0) \bar B_{\ell'' m'',\ell'm'}(q_0,q_0;q_0^2)\, ,\label{eq:BRH}
  \en
  where
  \eq\label{eq:H}
  H_{\ell m,\ell'm'}(q_0)=4\pi\sum_k\frac{v_{\ell m}^k\left(v_{\ell'm'}^k\right)^*}
  {q_k^2-q_0^2}
=
\frac{4\pi}{L^6}\,\sum_{{\bf p},{\bf q}}\mathscr{Y}^*_{\ell m}({\bf p})
\langle{\bf p}|G_L(q_0^2)|{\bf q}\rangle\mathscr{Y}_{\ell'm'}({\bf q})\, . 
\en
The modified quantization condition, derived from Eq.~(\ref{eq:BRH}) takes the form 
$\det\mathscr{A}=0$. Using again, as in Eq.~(\ref{eq:BKG}), the definition $K_\ell^M(q_0^2)=\left[R_\ell(q_0,q_0;q_0^2)\right]^{-1}$, we arrive at the following
expression for the matrix $\mathscr{A}_{\ell m,\ell'm'}$:
\eq\label{eq:quantization}
  \mathscr{A}_{\ell m,\ell'm'}(q_0)=\delta_{\ell \ell'}\delta_{mm'}K_\ell^M(q_0^2)-H_{\ell m,\ell' m'}(q_0)\, .
  \en

  \subsection{Calculation of the function $H_{\ell m,\ell' m'}(q_0)$}

  Owing to our choice of the superregular long-range potential, the quantity
  $H_{\ell m,\ell' m'}(q_0)$ is free of the ultraviolet divergences for $\ell,\ell'\leq \ell_{\sf max}$.
  However, one still needs a finite renormalization, in order to ensure that the definition
  of the function $H_{\ell m,\ell' m'}(q_0)$ in a finite volume is consistent with its
  infinite-volume counterpart. Below, we shall consider the cases $q_0^2<0$ and
  $q_0^2>0$ separately.

  \subsubsection{Negative energies, $q_0^2<0$}

  In the case $q_0^2<0$, a consistent definition of the loop function is given by
  \eq
  H_{\ell m,\ell' m'}(q_0)=(H_{\ell m,\ell' m'}(q_0)-H^\infty_{\ell m,\ell' m'}(q_0))
  +\frac{1}{4\pi}\,\delta_{\ell\ell'}\delta_{mm'}M_\ell(q_0)\, .
  \en
  It should be mentioned here that the functions $f_\ell(p,r)$ and hence $M_\ell(p)$ are
  analytic in the upper half of the complex $p$-plane.\footnote{Considering the Born series of the Green function $G_L(q_0^2)$, it is easy to get convinced that $M_\ell(q_0)$ is real everywhere below threshold.} Consequently, $M_\ell(q_0)$ is well-defined
  for negative values of $q_0^2$, taking into account the presence of the
  infinitesimal positive imaginary part in $q_0^2$. Furthermore,
  using Eqs.~(\ref{eq:longrangeonly}), (\ref{eq:H}) and applying the Poisson formula, one gets:
\eq
H_{\ell m,\ell' m'}(q_0)-H^\infty_{\ell m,\ell' m'}(q_0)=H^{(1)}_{\ell m,\ell' m'}(q_0)+H^{(2)}_{\ell m,\ell' m'}(q_0)+H^{(3)}_{\ell m,\ell' m'}(q_0)\, ,
\en
where
\eq\label{eq:H123}
H^{(1)}_{\ell m,\ell' m'}(q_0)&=&4\pi\sum_{{\bf n}}\int\frac{d^3{\bf p}}{(2\pi)^3}\,
\frac{\mathscr{Y}^*_{\ell m}({\bf p})\bigl(e^{i{\bf p}{\bf n}L}-1\bigr)\mathscr{Y}_{\ell'm'}({\bf p})}{{\bf p}^2-q_0^2}\, ,
\nonumber\\[2mm]
   H^{(2)}_{\ell m,\ell' m'}(q_0)&=&
  4\pi 
  \sum_{{\bf n},{\bf s}}\int\frac{d^3{\bf p}}{(2\pi)^3}\,\frac{d^3{\bf q}}{(2\pi)^3}\,
  \bigl(e^{i{\bf n}{\bf p}L-i{\bf s}{\bf q}L}-1\bigr)
  \frac{\mathscr{Y}^*_{\ell m}({\bf p})}{{\bf p}^2-q_0^2}\,
  \langle {\bf p}|T_L(q_0^2)|{\bf q}\rangle
 \frac{\mathscr{Y}_{\ell' m'}({\bf q})}{{\bf q}^2-q_0^2}\, ,
 \nonumber\\[2mm]
  H^{(3)}_{\ell m,\ell' m'}(q_0)&=&
  4\pi 
  \int\frac{d^3{\bf p}}{(2\pi)^3}\,\frac{d^3{\bf q}}{(2\pi)^3}\,
    \frac{\mathscr{Y}^*_{\ell m}({\bf p})}{{\bf p}^2-q_0^2}\,
 \biggl( \langle {\bf p}|T_L(q_0^2)|{\bf q}\rangle -\langle {\bf p}|T^\infty_L(q_0^2)|{\bf q}\rangle\biggr)
 \frac{\mathscr{Y}_{\ell' m'}({\bf q})}{{\bf q}^2-q_0^2}\, .\quad\quad\quad
 \en
 The first term here is the standard L\"uscher zeta-function. In order to calculate the
 remaining two terms, let us consider the LS equation for $T_L(q_0^2)$
 (a finite-volume counterpart of Eq.~(\ref{eq:longrangeonly})). Carrying out the partial-wave expansion
 \eq
 \langle {\bf p}|V_L|{\bf q}\rangle&=&4\pi\sum_{\ell m}Y_{\ell m}(\hat p)V_L^\ell(p,q)Y^*_{\ell m}(\hat q)\, ,
 \nonumber\\[2mm]
 \langle {\bf p}|T_L(q_0^2)|{\bf q}\rangle&=&4\pi\sum_{\ell m,\ell'm'}Y_{\ell m}
 (\hat p)T_L^{\ell m,\ell'm'}(p,q;q_0^2)Y^*_{\ell' m'}(\hat q)\, ,
 \en
 one gets
 \eq\label{eq:LSVV}
 &&T_L^{\ell m,\ell'm'}(p,q;q_0^2)=\delta^{\ell\ell'}\delta^{mm'}V_L^\ell(p,q)
\nonumber\\[2mm]
 &+&4\pi\sum_{\ell''m'}\int_0^\infty\frac{k^2dk}{(2\pi)^3}\,
 V_L^\ell(p,k)\frac{f_{\ell m,\ell''m''}(k)+f_{\ell m,\ell''m''}(-k)}{k^2-q_0^2}\, 
 T_L^{\ell'' m'',\ell'm'}(p,q;q_0^2)\, ,
 \en
 where
 \eq
f_{\ell m,\ell''m''}(k)+f_{\ell m,\ell''m''}(-k)
= \int d\Omega\, Y^*_{\ell m}(\hat k)\sum_{\bf n}e^{-i{\bf k}{\bf n}L}Y_{\ell''m''}(\hat k)\, .
  \en
  The quantity $f_{\ell m,\ell''m''}(k)$ is analytic in the upper half-plane of the variable $k$
  and vanishes exponentially, when ${\rm Im}\,k\to+\infty$. Note that $\ell+\ell'$ is always even for identical particles.
 
For the following discussion, it is convenient to define $V_L^\ell(p,q)$ for negative arguments,
 \eq
 V_L^\ell(p,q)=(-1)^\ell V_L^\ell(-p,q)=(-1)^\ell V_L^\ell(p,-q)=V_L^\ell(-p,-q)\, ,
 \en
 and, hence, from Eq.~(\ref{eq:LSVV}) one concludes that
 \eq
 T_L^{\ell m,\ell'm'}(p,q;q_0^2)&=&(-1)^\ell T_L^{\ell m,\ell'm'}(-p,q;q_0^2)=(-1)^{\ell'}T_L^{\ell m,\ell'm'}(p,-q;q_0^2)
 \nonumber\\[2mm]
 &=&T_L^{\ell m,\ell'm'}(-p,-q;q_0^2)\, .
  \en
 This means that the integration over the variable $k$ can be extended over the whole real axis, from $-\infty$ to $+\infty$:
 \eq\label{eq:LSV}
&& T_L^{\ell m,\ell'm'}(p,q;q_0^2)=\delta^{\ell\ell'}\delta^{mm'}V_L^\ell(p,q)
\nonumber\\[2mm]
 &+&4\pi\sum_{\ell''m'}\int_{-\infty}^\infty\frac{k^2dk}{(2\pi)^3}\,
 V_L^\ell(p,k)\frac{f_{\ell m,\ell''m''}(k)}{k^2-q_0^2}\, 
 T_L^{\ell'' m'',\ell'm'}(p,q;q_0^2)\, ,
 \en
 Using now the fact that $f_{\ell m,\ell''m''}(k)$ is analytic in the upper half-plane, one may
 shift the variables $p,q,k\to p,q,k+i\sigma$. The value of $\sigma$ is restricted by
 the singularities appearing in the free Green function as well as in the potential
 $V_L^\ell(p,q)$. Namely, $\sigma$ must fulfill the condition $\sigma<|q_0|$, for
 the free Green function to stay regular. The restriction coming from the potential
 does not depend on $q_0$. For instance, in case of Yukawa interaction,
 we have $\sigma<M_\pi/2$.
 The quantity $\sigma$ should obey both conditions.
 Performing the contour shift in
 the second and third lines of Eq.~(\ref{eq:H123}) as well, one sees that the
 finite-volume corrections to $H_{\ell m,\ell'm'}(q_0)$ are suppressed by the factor
 $e^{-\sigma L}$.

  \subsubsection{Positive energies, $q_0^2>0$}

  For $q_0^2>0$, the Poisson formula can not be used.
  The infinite-volume limit of  the quantity $H_{\ell m,\ell m}(q_0)$ in this case implies
  using the principal value prescription. Furthermore, from unitarity one can straightforwardly conclude that
  \eq
  ((2\ell+1)!!)^2 \langle G_L^\ell(q_0)\rangle&=& ((2\ell+1)!!)^2
  \langle G_L^\ell(q_0)\rangle_{\sf p.v.}
\nonumber\\[2mm]
  &+&i\frac{q_0^{2\ell+1}}{4\pi}\,\frac{(1+F^\ell(q_0))^2}{1-iq_0R_L^\ell(q_0,q_0;q_0^2)/(4\pi)}\, ,
    \en
    where
    \eq
    F_\ell(q_0)=\frac{1}{q_0^\ell}\,\,{\sf p.v.}\int_0^\infty\frac{p^2dp}{2\pi^2}\,\frac{p^\ell
      R_L^\ell(p,q_0;q_0^2)}    {p^2-q_0^2}\, .
    \en
    Here, $R_L^\ell$ denotes the $K$-matrix for the scattering on the long-range potential.
    Furthermore, since, by definition, ${\rm Im}\,\langle G_L^\ell(q_0)\rangle_{\sf p.v.}=0$,
    with the use of Eqs.~(\ref{eq:KlM}) and (\ref{eq:GLell}) one obtains, on the one hand,
    \eq\label{eq:1hand}
       {\rm Im}\,\langle G_L^\ell(q_0)\rangle
       =\frac{q_0^{2\ell+1}}{4\pi((2\ell+1)!!)^2}\,\frac{1}{|f_\ell(q_0)|^2}\, ,
       \en
       and, on the other hand,
\eq\label{eq:2hand}
   {\rm Im}\,\langle G_L^\ell(q_0)\rangle&=&\frac{q_0^{2\ell+1}}{4\pi((2\ell+1)!!)^2}\,\frac{(1+F^\ell(q_0))^2}{1+q_0^2R_L^\ell(q_0,q_0;q_0^2)/(4\pi)^2}
   \nonumber\\[2mm]
   &=&\frac{q_0^{2\ell+1}}{4\pi((2\ell+1)!!)^2}\,
   \left|1+\frac{1}{q_0^\ell}\,\int_0^\infty\frac{p^2dp}{2\pi^2}\,
   \frac{p^\ell T_L^\ell(p,q_0;q_0^2)}
               {p^2-q_0^2-i\varepsilon}\right|^2\, .
               \en
Using now Eq.~(\ref{eq:onshellpsi}) and performing the partial-wave expansion of the
               on-shell function in analogy with Eq.~(\ref{eq:Fuda-Whiting}), we obtain:
               \eq
               \frac{\phi_\ell(q_0,r)}{q_0r}=j_\ell(q_0r)+\int_0^\infty\frac{p^2dp}{2\pi^2}\,
               \frac{j_\ell(pr)T_L^\ell(p,q_0;q_0^2)}{p^2-q_0^2-i\varepsilon}\, ,
               \en
where $\phi_\ell(q_0,r)$ is the on-shell wave function. Performing the limit $r\to 0$ in this equation, we get:
\eq
\lim_{r\to 0}(2\ell+1)!!\frac{\phi_\ell(q_0r)}{(q_0r)^{\ell+1}}
=1+\frac{1}{q_0^\ell}\,\int_0^\infty\frac{p^2dp}{2\pi^2}\,
\frac{p^\ell T_L^\ell(p,q_0;q_0^2)}{p^2-q_0^2-i\varepsilon}\, .
\en
Finally, from Eq.~(\ref{eq:Newton}),
one concludes that Eqs.~(\ref{eq:1hand}) and (\ref{eq:2hand}) are consistent.
This represents a nice check of our approach.

A consistent definition of  the quantity $H_{\ell m,\ell' m'}(q_0)$ is given by
  \eq
  H_{\ell m,\ell' m'}(q_0)=(H_{\ell m,\ell' m'}(q_0)-H^\infty_{\ell m,\ell' m'}(q_0))
  +\delta_{\ell\ell'}\delta_{mm'}((2\ell+1)!!)^2 \langle G_L^\ell(q_0)\rangle_{\sf p.v.}\, ,
  \en
  where $H^\infty_{\ell m,\ell' m'}(q_0)$ is defined by a counterpart of Eq.~(\ref{eq:H}), with
  sums replaced by integrals with the principal-value prescription everywhere. Note also that, since the potential is superregular, no ultraviolet divergences arise except in the free loop containing no potential exchange. There, it can be handled, as usual,
  by using dimensional regularization and cancels
  anyway in the difference of the finite-volume and the infinite-volume contributions.

  Neglecting exponentially suppressed contributions from the long-range interactions,
  one could reduce the calculation of the function  $H_{\ell m,\ell' m'}(q_0)$ to the solution
  of the system of linear equations in the angular-momentum basis. This equation has the following
  form:
    \eq\label{eq:HH0}
  H_{\ell m,\ell'm'}(q_0)=H^0_{\ell m,\ell'm'}(q_0)
  +\sum_{\ell m,\ell''m''}H^0_{\ell m,\ell''m''}(q_0)q_0^{2\ell}R_L^{\ell''}(q_0) H_{\ell''m'',\ell'm'}(q_0)\, .
  \en
  Here, $R_L^\ell(q_0)=4\pi\tan\sigma_\ell(q_0)/q_0$ are the partial-wave on-shell
  $K$-matrices,
  corresponding to the long-range potential, and
  \eq
  H^0_{\ell m,\ell'm'}(q_0)=\frac{4\pi}{L^3}\,\sum_{\bf k}\frac{\mathscr{Y}_{\ell m}^*({\bf k})
  \mathscr{Y}_{\ell'm'}({\bf k})}{{\bf k}^2-q_0^2}\, .
  \en
  This quantity can be expressed through a linear combination of the L\"uscher zeta-functions. The quantity $H^\infty_{\ell m,\ell'm'}(q_0)$ should be taken equal to zero in this case.

Note however that neglecting exponential corrections coming from the long-range potential
  might be dangerous. This is seen, for example, from the fact that,
  below threshold, $q_0^2<0$, the quantity $R_L^\ell(q_0)$ develops the $t$-channel
  singularity that was mentioned earlier, whereas the exact function
  $H_{\ell m,\ell'm'}(q_0)$ is of course regular there.
The derivation of the modified
  quantization condition, which was presented above, nicely demonstrates the origin
  of the problem and a way to circumvent it. In fact, the problem is handmade and is
  not present in Eq.~(\ref{eq:quantization}). It emerges first, when one tries to evaluate
  $H_{\ell m,\ell'm'}(q_0)$ from Eq.~(\ref{eq:HH0}) and continue analytically below
  threshold. All this is perfectly consistent with the discussion in the recent
  paper~\cite{Raposo:2023oru}.

\subsection{Partial-wave mixing}

Above threshold, the modified zeta-function is determined from Eq.~(\ref{eq:HH0}) or
from the pertinent equation in the plane-wave basis.
Since $V_L$ is a long-range potential, it is expected that many partial waves will contribute
to this expression. However, this is not a problem, since $V_L$ is a well-known function,
with parameters that are determined very precisely elsewhere (e.g., the pion mass
and the pion axial-vector coupling, in case of the one-pion exchange potential).
Hence, the solution of Eq.~(\ref{eq:HH0}) does not require a fit to lattice data.
On the other hand, the short-range interaction, encoded in the function
$K_\ell^M(q_0^2)$, is determined from the fit. One expects that the partial-wave
mixing effect in the modified L\"uscher equation is small, exactly because of the
short-range nature of these interactions.

\subsection{Exponentially suppressed effects}

Up to now, we have consistently dropped the exponentially suppressed effects. However,
as mentioned already in the introduction, these effects can turn relatively large, owing
to the small mass scale. In the case of, say, $NN$ scattering, one may indirectly estimate
the size of the exponential effects, comparing the finite-volume spectra obtained in
the plane-wave basis with the solutions of the modified L\"uscher equation with the
same input. A simpler method to estimate the size of the exponential effects is the
comparison of the modified L\"uscher functions  $H_{\ell m,\ell'm'}(q_0)$, calculated
in the plane-wave basis and in the angular-momentum basis. This comparison does
not involve any parameters that characterize the short-range interactions.

There is one place, however, where one already knows that the exponential effects are
important. We remind the reader that the energy levels, which lie on the $t$-channel cut,
are indeed observed in the $NN$ system on the lattice~\cite{Green:2021qol}. Physical bound states cannot
be present there and, hence, the infinite-volume limit of the L\"uscher equation does not
predict a pole in this region.
The observed poles can only emerge because of the exponential contributions.

{
\subsection{The case with the short-range potential only}
The limit $V_L(r)\to 0$ is trivial. In this limit, the Jost function $f_L(q)=1$, $\sigma_\ell(q)=0$, and
Eq.~(\ref{eq:KlM}) reduces to the familiar expression
\eq
K_\ell^M(q^2)=q^{2\ell+1}\cot\delta_\ell(q)\, .
\en
Furthermore, the modified L\"uscher zeta-function $H_{\ell m,\ell'm'}$ reduces to the
conventional one, $Z_{\ell m,\ell'm'}$, and Eq.~(\ref{eq:quantization}) turns into the original L\"uscher equation.
}

\section{Comparison with the existing approaches}
\label{sec:comparison}

So far, several
different frameworks (including the present one) have been proposed to treat the finite-volume scattering in the presence of the long-range forces. All three approaches have one thing in common -- namely, they all treat the long-range part of the potential explicitly, without trying to approximate it by a string of contact interactions (like in the derivation of the ordinary L\"uscher equation). After that point, the paths start to diverge.

In the recent paper~\cite{Raposo:2023oru} a modified two-body quantization condition
has been derived in the presence of both the long- and short-range forces. The authors
present their central result in two different forms. Namely, Eq.~(3.63) of that paper
is written down in a plane wave {\em and} angular-momentum basis. From this point
of view, it bears strong resemblance with the approach of Ref.~\cite{Meng:2021uhz},
however, with a conceptual difference. Namely, all short-range interactions
in Ref.~\cite{Raposo:2023oru} are summed up and enter the quantization
condition through an auxiliary
{\em on-shell} $K$-matrix $\bar K^{\sf os}$.
No particular parameterization of the {\em on-shell} $K$-matrix $\bar K^{\sf os}$ is specified. We note, however, that for any realistic application to lattice data involving partial-wave mixing such a parameterization would be required.
In contrast,  the long- and
short-range interactions are treated on equal footing in Ref.~\cite{Meng:2021uhz}, and the short-range $K$-matrix
is implicitly parameterized in terms of the effective couplings appearing in the
Hamiltonian.

Furthermore, in Eqs.~(5.3) and (5.4) of Ref.~\cite{Raposo:2023oru} the authors recast
their central result in the angular-momentum basis. This result bears close analogy to
our modified quantization condition. For example, the quantity $F^T$ from their
Eq.~(5.4) is similar to our modified L\"uscher function $H_{\ell m,\ell'm'}(q_0)$. The main
difference, as already noted in the introduction, is that the authors of
Ref.~\cite{Raposo:2023oru} propose a two-step procedure for the analysis of lattice
data. Namely, at the first step, an auxiliary matrix $\bar K^{\sf os}$ is determined
from data. At the next step, $\bar K^{\sf os}$ is substituted into the integral equation
which is solved to obtain the physical $K$-matrix. We propose to unite these two steps
in one -- in our approach, the auxiliary $K$-matrix is related
to the physical one at the same CM energy through a simple algebraic expression.

Last but not least, it has been recently proposed to solve the $t$-channel problem in the two-body scattering
by writing down three-body scattering equations~\cite{Hansen:2024ffk}. In particular, the $t$-channel cut
that emerges close to threshold in $DD^*$ scattering (assuming a stable $D^*$) does not show up
in the three-body quantization condition for the $DD\pi$ system, even if the bound state in the
$D\pi$ subsystem lies below the elastic threshold. The results seems surprising at a first glance.
Let us recall however that the three-body quantization condition is written down in the space
of spectator momenta. Hence, in case of the stable $D^*$ meson, this approach, up to the exponentially suppressed contributions should be algebraically 
equivalent to the plane-wave solution proposed in Refs.~\cite{Meng:2021uhz,Du:2023hlu,Meng:2023bmz}. Note however
that, in difference to the latter, the approach of Ref.~\cite{Hansen:2024ffk} allows for a smooth
transition to the case of an unstable $D^*$ meson.\footnote{Note also that the calculation of the infinite-volume scattering amplitude in the region of the $t$-channel cut has been carried out earlier in Ref.~\cite{Dawid:2023jrj}. In these calculations it was explicitly shown that the particle-dimer amplitude develops the left-hand singularity. The method of Ref.~\cite{Hansen:2024ffk} essentially stems from this observation.}

  To summarize, the difference between the existing approaches mainly boils down to
  the following two points:
  \begin{enumerate}
  \item
    {\em Technical convenience.} The quantization condition can be written down
    in the plane-wave basis as well as in the angular-momentum basis. Given
    modern computing capacities, the difference between these two representations is not a decisive factor anymore. Despite this, we still prefer a more compact representation
    in the angular-momentum basis, which reduces to a single algebraic equation if the partial-wave mixing for the
    short-range interactions can be neglected.
    The same statement applies to relating physical observables to quantities
    extracted from the fit to lattice data. For example, in order to
    relate $\bar K^{\sf os}$ to the physical $K$-matrix, integral equations need to be
    solved~\cite{Raposo:2023oru}, whereas the corresponding  link in our approach
    is given by a simple algebraic relation.
  \item
    {\em The choice of quantities extracted from lattice data.}
    This is a more subtle issue. In an idealized world with single-channel
    scattering and no partial-wave mixing, the L\"uscher equation just gives the scattering
    phase in terms of the level energy. In realistic situations, however, one often needs a parameterization of the $K$-matrix in order to solve the
    quantization condition. Without any doubt, the use of an effective Hamiltonian provides such
    a parameterization within the range of applicability of this particular
    EFT. In this case, the quantities that are extracted from data at first hand
    are the couplings of the effective Hamiltonian. However, expressing
    the lattice energy levels directly in terms of the physical $K$-matrix, in our opinion,
    renders the approach more flexible: for example, one might use different EFTs
 (or expansions) in different energy regions to cover a larger energy range.
\end{enumerate}

\section{Conclusions}
\label{sec:concl}

\begin{itemize}

\item[i)]
  In this paper, we have derived a modified L\"uscher equation in the presence of
  both the long-range and short-range interactions. The presence of the former
  leads to several (interrelated) conceptual difficulties in the
  standard L\"uscher equation. Namely,
\begin{itemize}
  \item The partial-wave expansion may converge slowly, and hence there could be a significant admixture of the higher partial
    waves in the L\"uscher equation that complicates the analysis of data.
  \item
    The long-range interactions lead to a $t$-channel cut in the scattering amplitude
    that moves very close to the threshold, if the range of the interactions increases.
    Using lattice energy levels that lie below the $t$-channel threshold in the L\"uscher
    equation is inconsistent.
  \item
    The exponentially suppressed contributions could be still significant for not so large
    values of $L$.
\end{itemize}
Our approach which, loosely speaking, represents a re-formulation of the modified
effective-range expansion of Ref.~\cite{vanHaeringen:1981pb} in a finite volume,
is capable to address all above challenges.

\item[ii)]
   Several alternative approaches have appeared recently in the literature~\cite{Meng:2021uhz,Raposo:2023oru,Hansen:2024ffk}. In our paper, a detailed comparison to these approaches is given. We argue that
  our method is conceptually closest to the original L\"uscher framework. 
  It allows to directly extract the scattering phase shift from the measured energy spectrum, if the partial-wave mixing for the short-range interactions is negligible and if the parameters of the long-range potential (i.e., the mass and the coupling of the pion) are known accurately for a given lattice ensemble.  
  Hence, it should be possible to analyze lattice data analog to the original L\"uscher approach
  once the modified L\"uscher function is available.

\item[iii)]
  The modified L\"uscher function, which incorporates the long-range interaction, is
  a central ingredient of our approach. In the present paper we consider the evaluation
  of this function in great detail, paying particular attention to the issues of the ultraviolet
  divergences and renormalization. Once this function, which does not depend on the
  unknown parameters of the short-range force, is calculated and tabulated, the
  analysis of data exactly follows the standard pattern. An explicit calculation of this
  function is however a rather challenging enterprise and will be discussed in a
  separate publication.

\item[iv)]
  Note that in this paper we deliberately ignored all issues related with the spin of particles, moving frames, relativistic effects, etc. All this is inessential in the context of
  the problems considered here and would only blur the discussion.

\item[v)]
  It remains to be seen, whether the Coulomb interaction can be treated consistently
  in the same manner, and whether the results would add something substantial
  to the findings of Refs.~\cite{Beane:2014qha,Cai:2018why,Christ:2021guf}. Here, it should be also mentioned
  that, due to the removal of the zero mode of the Coulomb field in case of periodic
  boundary conditions, the resulting Lagrangian is not local anymore. This, in its turn,
  might cause problems in the matching of the non-relativistic effective field theory,
  which is used
  for the derivation of the L\"uscher equation, to its relativistic counterpart
  (see, e.g.,~Ref.~\cite{Davoudi:2018qpl}). In this context, it would be interesting to 
  explore the possibility of using different boundary conditions. An alternative to this
  would be to use the formulation with massive protons, see, e.g.~\cite{Endres:2015gda}.
  
\item[vi)]
  The major challenge consists in using the same method in the three-particle problem. 
  For instance, it remains to be seen, whether the long-range one-pion exchange force
  in the three-nucleon system can be separated as neatly from the short-range
  interactions as done in case of the nucleon-nucleon scattering.

\end{itemize}

\begin{sloppypar}
{\em Acknowledgments:} The authors thank Vadim Baru, Evgeny Epelbaum, Jambul Gegelia,
Christoph Hanhart, Nils Hermanssohn-Truedsson, Bai-Long Hoid, Lu Meng and Fernando Romero-Lopez for interesting discussions.
	The work of R.B, F.M. and A.R. was  funded in part by
	the Deutsche Forschungsgemeinschaft
	(DFG, German Research Foundation) – Project-ID 196253076 – TRR 110 and by the Ministry of Culture and Science of North Rhine-Westphalia through the NRW-FAIR project.
	A.R., in addition, thanks Volkswagenstiftung 
	(grant no. 93562) and the Chinese Academy of Sciences (CAS) President's
	International Fellowship Initiative (PIFI) (grant no. 2024VMB0001) for the
	partial financial support.
	The work of J.-Y.P. and J.-J.W. was supported by the National Natural Science Foundation of China (NSFC) under Grants No. 12135011, 12175239, 12221005, and by the National Key R\&D Program of China under Contract No. 2020YFA0406400, and by the Chinese Academy of Sciences under Grant No. YSBR-101, and by the Xiaomi Foundation / Xiaomi Young Talents Program.
        H.-W.H. was supported by Deutsche Forschungsgemeinschaft
  (DFG, German Research Foundati  on) --
  Project ID 279384907 -- SFB 1245.
\end{sloppypar}

\appendix
  
\section{Calculating the Green function}

\label{sec:calcGreen}
    
 The Green function in the coordinate space can
    be expressed through the M\o{}ller operator
\eq\label{eq:GL-coordinate}
 \langle {\bf r}|G_L(q_0^2+i\varepsilon)|{\bf w}\rangle
 =\int\frac{d^3{\bf p}}{(2\pi)^3}\,
 \langle {\bf r}|\Omega(q_0^2+i\varepsilon)|{\bf p}\rangle
 \frac{e^{-i{\bf p}{\bf w}}}{{\bf p}^2-q_0^2-i\varepsilon}\, .
 \en
 In Ref.~\cite{Fuda:1973zz}, Fuda and Whiting defined the off-shell
 scattering wave function
 (cf. with Eq.~(\ref{eq:onshellpsi})):
 \eq\label{eq:Fuda-Whiting}
 \langle {\bf r}|\Omega(q_0^2+i\varepsilon)|{\bf p}\rangle=
4\pi\sum_{\ell m}
Y_{\ell m}(\hat r)i^\ell\frac{\phi_\ell(q_0,p,r)}{pr}\,
Y_{\ell m}^*(\hat p)\, .
\en
This scattering wave function obeys the equation
\eq\label{eq:A3}
\biggl(q_0^2+\frac{d^2}{dr^2}-\frac{\ell(\ell+1)}{r^2}-V_L(r)\biggr)
\phi_\ell(q_0,p,r)=(q_0^2-p^2)u_\ell(pr)\, ,
\en
where
\eq
u_\ell(z)=zj_\ell(z)\, ,\quad\quad
v_\ell(z)=zn_\ell(z)\, ,\quad\quad
w_\ell^{(\pm)}(z)=zh^{(\pm)}(z)=-v_\ell(z)\pm iu_\ell(z)
\en
are expressed through the spherical Bessel, Neumann and Hankel functions, respectively. The familiar on-shell wave function is given by
$\phi_\ell(p,r)\doteq\phi_\ell(p,p,r)$.

Using the expansion of the plane wave into spherical functions in Eq.~(\ref{eq:GL-coordinate}), we obtain
\eq
G_L^\ell(r,w;q_0^2+i\varepsilon)=
 4\pi\int_0^\infty\frac{p^2dp}{(2\pi)^3}\,
\frac{\phi_\ell(q_0,p,r)}{pr^{\ell+1}}\,
\frac{j_\ell(pw)}{w^\ell}\,\frac{1}{p^2-q_0^2-i\varepsilon}\, .
\en
Performing the limit $w\to 0$, one gets:
\eq\label{eq:GellL}
G_L^\ell(r,0;q_0^2+i\varepsilon)=
\frac{4\pi}{(2\ell+1)!!}\,\frac{(2i)^\ell \ell!}{(2\ell+1)!}\,
\int_0^\infty\frac{pdp}{(2\pi)^3}\,
\frac{D^{2\ell+1}\phi_\ell(q_0,p,r)}
{p^2-q_0^2-i\varepsilon}\, ,
\en
where
\eq\label{eq:D2ellplus1}
D^{2\ell+1}\phi_\ell(q_0,p,r)\doteq
\left(\frac{-ip}{2}\right)^\ell\frac{1}{\ell!}\frac{d^{2\ell+1}}{dr^{2\ell+1}}\,r^\ell \phi_\ell(q_0,p,r)\, .
\en
Note that the same definition of the operator $D^{2\ell+1}$ is used, if $\phi_\ell$ is
replaced by an arbitrary function.

In order to perform the integral over $p$, we rewrite the wave function
$\phi_\ell$ in terms of the off-shell functions $f_\ell$~\cite{Fuda:1973zz}:
\eq\label{eq:offshellphi}
\phi_\ell(q_0,p,r)&=&-\frac{\pi p}{2}\left(\frac{q_0}{p}\right)^\ell\frac{f_\ell(q_0,p)-f_\ell(q_0,-p)}{i\pi p f_\ell(q_0)}\,e^{-i\ell\pi/2}f_\ell(q_0,r)
\nonumber\\[2mm]
&+&\frac{1}{2i}\,\left(e^{-i\ell\pi/2}f_\ell(q_0,p,r)
-e^{i\ell\pi/2}f_\ell(q_0,-p,r)\right)\, ,
\en
where $f_\ell(q_0,r)\doteq f_\ell(q_0,q_0,r)$.
Here, the function $f_\ell$ obeys the equation
\eq\label{eq:equationfell}
\biggl(q_0^2+\frac{d^2}{dr^2}-\frac{\ell(\ell+1)}{r^2}-V_L(r)\biggr)
f_\ell(q_0,p,r)=(q_0^2-p^2)e^{i\ell\pi/2}w^{(+)}_\ell(pr)\, ,
\en
and has the asymptotic normalization
\eq
f_\ell(q_0,p,r)\sim e^{ipr}\, ,\quad\quad \mbox{as}~r\to\infty\, .
\en
The off-shell Jost functions are defined as
\eq
f_\ell(q_0,p)=\frac{p^\ell e^{-i\ell\pi/2}(2\ell+1)}{(2\ell+1)!!}\,
\lim_{r\to 0}r^\ell f_\ell(q_0,p,r)\, ,
\en
and the usual Jost functions are obtained from the off-shell Jost functions
according to $f_\ell(q_0)=f_\ell(q_0,q_0)$.

Substituting Eq.~(\ref{eq:offshellphi}) into Eq.~(\ref{eq:GellL}), it is seen
that the integration can be extended from $-\infty$ to $+\infty$, owing to the
symmetry of the integrand:
\eq
G_L^\ell(r,0;q_0^2+i\varepsilon)&=&
\frac{1}{(2\ell+1)!!}\,\frac{(2i)^\ell \ell!}{(2\ell+1)!}\,2\pi ie^{-i\ell\pi/2}
\int_{-\infty}^\infty\frac{pdp}{(2\pi)^3(p^2-q_0^2-i\varepsilon)}
\nonumber\\[2mm]
&\times&\left(
\frac {f_\ell(q_0,p)}{f_\ell(q_0)}\,
D^{2\ell+1}f_\ell(q_0,r)-D^{2\ell+1}f_\ell(q_0,p,r)\right)\, .
\en
Note that the factor $\left(q_0/p\right)^\ell$ has disappeared, since
$D^{2\ell+1}f_\ell(q_0,r)$ contains the factor $q_0^\ell$ instead of $p^\ell$,
cf. Eq.~(\ref{eq:D2ellplus1}).

In order to perform the integral by using Cauchy's theorem, it is important
to show that the Jost solutions do not have singularities in the upper complex
plane of the variable $p$. To this end, we define the functions
\eq
g_\ell(q_0,p,r)&=&\frac{e^{-i\ell\pi/2}(2\ell+1)}{(2\ell+1)!!}\,(pr)^\ell f_\ell(q_0,p,r)\, ,
\nonumber\\[2mm]
z_\ell(pr)&=&\frac{(2\ell+1)}{(2\ell+1)!!}\,(pr)^\ell w^{(+)}(pr)\, .
\en
Using Eq.~(\ref{eq:equationfell}) and the asymptotic condition, it can
be shown that the function $g_\ell$ obeys the following integral equation
\eq
g_\ell(q_0,p,r)&=&z_\ell(pr)-\frac{1}{q_0}\,\int_0^\infty dw\theta(w-r)
\left(\frac{r}{w}\right)^\ell\left(u_\ell(q_0r)v_\ell(q_0w)
-v_\ell(q_0r)u_\ell(q_0w)\right)
\nonumber\\[2mm]
&\times&V_L(w)g_\ell(q_0,p,w)\, .
\en
Solving this equation iteratively, one arrives at
\eq\label{eq:gell}
g_\ell(q_0,p,r)=z_\ell(pr)+\int_0^\infty dw K_\ell(r,w;q_0)z_\ell(pw)\, .
\en
An exact form of the kernel $K_\ell$ is not important.
It suffices to know that the kernel does not depend on $p$ and vanishes at $w<r$. Furthermore, assuming $r\to 0$, we get
\eq
f_\ell(q_0,p)=z_\ell(0)+\int_0^\infty dw K_\ell(0,w;q_0)z_\ell(pw)\, \, ,\quad\quad
z_\ell(0)=\frac{(2\ell+1)!}{l!2^\ell(2\ell+1)!!}\, .
\en
Acting now with the operator $D^{2\ell+1}$ on Eq.~(\ref{eq:gell}) and taking the limit $r\to 0$, one gets:
\eq
\lim_{r\to 0}D^{2\ell+1}f_\ell(q_0,p,r)=p^{2\ell+1}\tilde z_\ell(0)+p^\ell\int_0^\infty dw \tilde K_\ell(0,w;q_0)z_\ell(pw)\, .
\en
Again, $\tilde z_\ell$, $\tilde K_\ell$ are independent of $p$. Performing now
Cauchy integrals, one gets:
\eq\label{eq:zell0}
\int_{-\infty}^\infty\frac{pdp}{2\pi i}\,\frac{f_\ell(q_0,p)}{p^2-q_0^2-i\varepsilon}&=&z_\ell(0)\int_{-\infty}^\infty\frac{pdp}{2\pi i}\,\frac{1}{p^2-q_0^2-i\varepsilon}+\frac{1}{2}\,\int_0^\infty K_\ell(0,w;q_0)z_\ell(q_0w)
\nonumber\\[2mm]
&=&-\frac{1}{2}\,z_\ell(0)+\frac{1}{2}\,f_\ell(q_0)\, .
\en
Here, one has used the fact that the integral, multiplying $z_\ell(0)$, vanishes in the symmetric boundaries. Furthermore
\eq\label{eq:zell0t}
&&\lim_{r\to 0}\int_{-\infty}^\infty\frac{pdp}{2\pi i}\,\frac{D^{2\ell+1}f_\ell(q_0,p,r)}{p^2-q_0^2-i\varepsilon}
\nonumber\\[2mm]
&=&\tilde z_\ell(0)\int_{-\infty}^\infty\frac{pdp}{2\pi i}\,\frac{p^{2\ell+1}}{p^2-q_0^2-i\varepsilon}+\frac{q_0^\ell}{2}\,\int_0^\infty dw \tilde K_\ell(0,w;q_0)z_\ell(q_0w)
\nonumber\\[2mm]
&=&\tilde z_\ell(0)\int_{-\infty}^\infty\frac{dp(p^{2\ell+2}-q_0^{2\ell+2})}{p^2-q_0^2+i\varepsilon}+\frac{1}{2}\,\lim_{r\to 0}D^{2\ell+1}f_\ell(q_0,r)
\nonumber\\[2mm]
&=&\tilde z_\ell(0)X_\ell(q_0^2)
+\frac{1}{2}\,\lim_{r\to 0}D^{2\ell+1}f_\ell(q_0,r)\, .
\en
Here, $X_\ell(q_0^2)$ denotes a polynomial of order $\ell$ in the variable
$q_0^2$. The coefficients of this polynomial are ultraviolet-divergent and can
be regularized, e.g., introducing a momentum cutoff on the integration momenta,
$|p|\leq \Lambda$.
  
Collecting all factors together, we obtain
\eq
\langle G_L^\ell(q_0)\rangle=\frac{1}{4\pi\left((2\ell+1)!!\right)^2}\,M_\ell(q_0)+\mbox{real polynomial in $q_0^2$}\, ,
  \en
  where $M_\ell(q_0)$ is given by Eq.~(\ref{eq:Mell}).

   One more remark is in order. It should be pointed out that the final result crucially depends on the
  validity of Eqs.~(\ref{eq:zell0}) and (\ref{eq:zell0t}). Using Cauchy's
  theorem straightforwardly
  is not allowed, since the integrand does not vanish sufficiently fast at the infinity.
  The result given above corresponds to the choice of symmetric boundary conditions
  $-\Lambda\leq p\leq\Lambda$ and $\Lambda\to\infty$, which follows from extending
  the initial integration area by using the fact that the integrand is even under the interchange $p\leftrightarrow -p$.
  The terms containing the potential are vanishing exponentially on a large semicircle
in the complex plane, and so Cauchy's theorem can be used there without further ado.

\section{Cancellation of the poles}
\label{app:cancellation}

Using Eq.~(\ref{eq:twopotential}), it is straightforward to see that the full Green function $G=G_0+G_0TG_0$
can be espressed as
\eq\label{eq:G-GL}
G(q_0^2)=G_L(q_0^2)+G_L(q_0^2)T_S(q_0^2)G_L(q_0^2)\, .
\en
Our aim is to show that the poles of $G_L(q_0^2)$ will cancel in $G(q_0^2)$. Note that our reasoning will be valid both in a finite as well as in the infinite volume.

Owing to the spectral representation written down in Eq.~(\ref{eq:spectral}), in the vicinity of an isolated pole at $q_0^2=q_n^2$, the Green function $G_L$ has the following representation:
\eq\label{eq:split}
G_L(q_0^2)=\frac{|\psi_n\rangle\langle\psi_n|}{q_n^2-q_0^2}+\tilde G_L(q_0^2)\, ,
\en
where $\tilde G_L(q_0^2)$ is regular at $q_0^2=q_n^2$. Furthermore, defining the quantity
\eq
\tilde T_S(q_0^2)=V_S+V_S\tilde G_L(q_0^2)\tilde T_S(q_0^2)\, ,
\en
which is apparenly regular at $q_0^2=q_n^2$, we obtain
\eq\label{eq:TSnopole}
T_S(q_0^2)=\tilde T_S(q_0^2)+\frac{\tilde T_S(q_0^2)|\psi_n\rangle\langle\psi_n|\tilde T_S(q_0^2)}
{q_n^2-q_0^2-\langle\psi_n|\tilde T_S(q_0^2)|\psi_n\rangle}\, .
\en
It is explicitly seen that this expression does not contain a pole at $q_0^2=q_n^2$, if the matrix element in the denominator does not accidentally vanish. Moreover, using Eqs.~(\ref{eq:split}) and 
(\ref{eq:TSnopole}) in Eq.~(\ref{eq:G-GL}), after a simple algebra one obtains:
\eq
G=\tilde G_L+\tilde G_L\tilde T_S\tilde G_L
+\frac{(1+\tilde G_L\tilde T_S)|\psi_n\rangle
\langle\psi_n|(\tilde T_S\tilde G_L+1)}{q_n^2-q_0^2-\langle\psi_n|\tilde T_S(q_0^2)|\psi_n\rangle}\, .
\en
Again, the poles that emerge from $G_L$, have canceled in the final result.

\bibliographystyle{unsrt}
\bibliography{ref}

\begin{thebibliography}{10}

\bibitem{Luscher:1990ux}
Martin L{\"u}scher.
\newblock {Two particle states on a torus and their relation to the scattering
  matrix}.
\newblock {\em Nucl. Phys. B}, 354:531--578, 1991.

\bibitem{Rummukainen:1995vs}
K.~Rummukainen and Steven~A. Gottlieb.
\newblock {Resonance scattering phase shifts on a nonrest frame lattice}.
\newblock {\em Nucl. Phys. B}, 450:397--436, 1995.

\bibitem{Lage:2009zv}
Michael Lage, Ulf-G. Mei{\ss}ner, and Akaki Rusetsky.
\newblock {A Method to measure the antikaon-nucleon scattering length in
  lattice QCD}.
\newblock {\em Phys. Lett. B}, 681:439--443, 2009.

\bibitem{Bernard:2010fp}
V.~Bernard, M.~Lage, U.-G. Mei{\ss}ner, and A.~Rusetsky.
\newblock {Scalar mesons in a finite volume}.
\newblock {\em JHEP}, 01:019, 2011.

\bibitem{He:2005ey}
Song He, Xu~Feng, and Chuan Liu.
\newblock {Two particle states and the S-matrix elements in multi-channel
  scattering}.
\newblock {\em JHEP}, 07:011, 2005.

\bibitem{Liu:2005kr}
Chuan Liu, Xu~Feng, and Song He.
\newblock {Two particle states in a box and the S-matrix in multi-channel
  scattering}.
\newblock {\em Int. J. Mod. Phys. A}, 21:847--850, 2006.

\bibitem{Hansen:2012tf}
Maxwell~T. Hansen and Stephen~R. Sharpe.
\newblock {Multiple-channel generalization of Lellouch-Luscher formula}.
\newblock {\em Phys. Rev. D}, 86:016007, 2012.

\bibitem{Briceno:2012yi}
Raul~A. Briceno and Zohreh Davoudi.
\newblock {Moving multichannel systems in a finite volume with application to
  proton-proton fusion}.
\newblock {\em Phys. Rev. D}, 88(9):094507, 2013.

\bibitem{Li:2012bi}
Ning Li and Chuan Liu.
\newblock {Generalized L\"uscher formula in multichannel baryon-meson
  scattering}.
\newblock {\em Phys. Rev. D}, 87(1):014502, 2013.

\bibitem{Guo:2012hv}
Peng Guo, Jozef Dudek, Robert Edwards, and Adam~P. Szczepaniak.
\newblock {Coupled-channel scattering on a torus}.
\newblock {\em Phys. Rev. D}, 88(1):014501, 2013.

\bibitem{Leskovec:2012gb}
Luka Leskovec and Sasa Prelovsek.
\newblock {Scattering phase shifts for two particles of different mass and
  non-zero total momentum in lattice QCD}.
\newblock {\em Phys. Rev. D}, 85:114507, 2012.

\bibitem{Kim:2005gf}
C.~h. Kim, C.~T. Sachrajda, and Stephen~R. Sharpe.
\newblock {Finite-volume effects for two-hadron states in moving frames}.
\newblock {\em Nucl. Phys. B}, 727:218--243, 2005.

\bibitem{Gockeler:2012yj}
M.~Gockeler, R.~Horsley, M.~Lage, U.-G. Mei{\ss}ner, P.~E.~L. Rakow,
  A.~Rusetsky, G.~Schierholz, and J.~M. Zanotti.
\newblock {Scattering phases for meson and baryon resonances on general
  moving-frame lattices}.
\newblock {\em Phys. Rev. D}, 86:094513, 2012.

\bibitem{Meng:2021uhz}
Lu~Meng and E.~Epelbaum.
\newblock {Two-particle scattering from finite-volume quantization conditions
  using the plane wave basis}.
\newblock {\em JHEP}, 10:051, 2021.

\bibitem{Raposo:2023nex}
Andr\'e Bai\~ao Raposo and Maxwell~T. Hansen.
\newblock {The L\"uscher scattering formalism on the t-channel cut}.
\newblock {\em PoS}, LATTICE2022:051, 2023.

\bibitem{Raposo:2023oru}
Andr\'e Bai\~ao Raposo and Maxwell~T. Hansen.
\newblock {Finite-volume scattering on the left-hand cut}.
\newblock 11 2023.

\bibitem{Green:2021qol}
Jeremy~R. Green, Andrew~D. Hanlon, Parikshit~M. Junnarkar, and Hartmut Wittig.
\newblock {Weakly bound $H$ dibaryon from SU(3)-flavor-symmetric QCD}.
\newblock {\em Phys. Rev. Lett.}, 127(24):242003, 2021.

\bibitem{Du:2023hlu}
Meng-Lin Du, Arseniy Filin, Vadim Baru, Xiang-Kun Dong, Evgeny Epelbaum,
  Feng-Kun Guo, Christoph Hanhart, Alexey Nefediev, Juan Nieves, and Qian Wang.
\newblock {Role of Left-Hand Cut Contributions on Pole Extractions from Lattice
  Data: Case Study for Tcc(3875)+}.
\newblock {\em Phys. Rev. Lett.}, 131(13):131903, 2023.

\bibitem{Meng:2023bmz}
Lu~Meng, Vadim Baru, Evgeny Epelbaum, Arseniy~A. Filin, and Ashot~M. Gasparyan.
\newblock {Solving the left-hand cut problem in lattice QCD: $T_{cc}(3875)^+$
  from finite volume energy levels}.
\newblock 12 2023.

\bibitem{Baru:2015ira}
V.~Baru, E.~Epelbaum, A.~A. Filin, and J.~Gegelia.
\newblock {Low-energy theorems for nucleon-nucleon scattering at unphysical
  pion masses}.
\newblock {\em Phys. Rev. C}, 92(1):014001, 2015.

\bibitem{Baru:2016evv}
V.~Baru, E.~Epelbaum, and A.~A. Filin.
\newblock {Low-energy theorems for nucleon-nucleon scattering at $M_\pi=450$
  MeV}.
\newblock {\em Phys. Rev. C}, 94(1):014001, 2016.

\bibitem{Hansen:2014eka}
Maxwell~T. Hansen and Stephen~R. Sharpe.
\newblock {Relativistic, model-independent, three-particle quantization
  condition}.
\newblock {\em Phys. Rev.}, D90(11):116003, 2014.

\bibitem{Hansen:2015zga}
Maxwell~T. Hansen and Stephen~R. Sharpe.
\newblock {Expressing the three-particle finite-volume spectrum in terms of the
  three-to-three scattering amplitude}.
\newblock {\em Phys. Rev.}, D92(11):114509, 2015.

\bibitem{Hammer:2017uqm}
Hans-Werner Hammer, Jin-Yi Pang, and A.~Rusetsky.
\newblock {Three-particle quantization condition in a finite volume: 1. The
  role of the three-particle force}.
\newblock {\em JHEP}, 09:109, 2017.

\bibitem{Hammer:2017kms}
H.~W. Hammer, J.~Y. Pang, and A.~Rusetsky.
\newblock {Three particle quantization condition in a finite volume: 2. General
  formalism and the analysis of data}.
\newblock {\em JHEP}, 10:115, 2017.

\bibitem{Mai:2017bge}
M.~Mai and M.~{D\"{o}ring}.
\newblock {Three-body Unitarity in the Finite Volume}.
\newblock {\em Eur. Phys. J.}, A53(12):240, 2017.

\bibitem{Mai:2018djl}
Maxim Mai and Michael D{\"{o}}ring.
\newblock {Finite-Volume Spectrum of $\pi^+\pi^+$ and $\pi^+\pi^+\pi^+$
  Systems}.
\newblock {\em Phys. Rev. Lett.}, 122(6):062503, 2019.

\bibitem{Hansen:2024ffk}
Maxwell~T. Hansen, Fernando Romero-L\'opez, and Stephen~R. Sharpe.
\newblock {Incorporating $DD\pi$ effects and left-hand cuts in lattice QCD
  studies of the $T_{cc}(3875)^+$}.
\newblock 1 2024.

\bibitem{Beane:2014qha}
Silas~R. Beane and Martin~J. Savage.
\newblock {Two-Particle Elastic Scattering in a Finite Volume Including QED}.
\newblock {\em Phys. Rev. D}, 90(7):074511, 2014.

\bibitem{Cai:2018why}
Yiming Cai and Zohreh Davoudi.
\newblock {QED-corrected Lellouch-Luescher formula for $K \rightarrow \pi\pi$
  decay}.
\newblock {\em PoS}, LATTICE2018:280, 2018.

\bibitem{Christ:2021guf}
Norman Christ, Xu~Feng, Joseph Karpie, and Tuan Nguyen.
\newblock {\ensuremath{\pi}-\ensuremath{\pi} scattering, QED, and finite-volume
  quantization}.
\newblock {\em Phys. Rev. D}, 106(1):014508, 2022.

\bibitem{vanHaeringen:1981pb}
H.~van Haeringen and L.~P. Kok.
\newblock {Modified Effective Range Function}.
\newblock {\em Phys. Rev. A}, 26:1218--1225, 1982.

\bibitem{Kong:1999sf}
Xinwei Kong and Finn Ravndal.
\newblock {Coulomb effects in low-energy proton proton scattering}.
\newblock {\em Nucl. Phys. A}, 665:137--163, 2000.

\bibitem{Badalian:1981xj}
A.~M. Badalian, L.~P. Kok, M.~I. Polikarpov, and Yu.~A. Simonov.
\newblock {Resonances in Coupled Channels in Nuclear and Particle Physics}.
\newblock {\em Phys. Rept.}, 82:31--177, 1982.

\bibitem{Goldberger-Watson}
Marvin~L. Goldberger and Kenneth~M. Watson.
\newblock {\em {Collision Theory}}.
\newblock Dover Publications, 10 2004.

\bibitem{Birse:1998dk}
Michael~C. Birse, Judith~A. McGovern, and Keith~G. Richardson.
\newblock {A Renormalization group treatment of two-body scattering}.
\newblock {\em Phys. Lett. B}, 464:169--176, 1999.

\bibitem{Steele:1998zc}
James~V. Steele and R.~J. Furnstahl.
\newblock {Removing pions from two nucleon effective field theory}.
\newblock {\em Nucl. Phys. A}, 645:439--461, 1999.

\bibitem{Doring:2009bi}
M.~Doring, C.~Hanhart, F.~Huang, S.~Krewald, and U.~G. Meissner.
\newblock {The Role of the background in the extraction of resonance
  contributions from meson-baryon scattering}.
\newblock {\em Phys. Lett. B}, 681:26--31, 2009.

\bibitem{Dawid:2023jrj}
Sebastian~M. Dawid, Md~Habib~E. Islam, and Ra\'ul~A. Brice\~no.
\newblock {Analytic continuation of the relativistic three-particle scattering
  amplitudes}.
\newblock {\em Phys. Rev. D}, 108(3):034016, 2023.

\bibitem{Davoudi:2018qpl}
Zohreh Davoudi, James Harrison, Andreas J\"uttner, Antonin Portelli, and
  Martin~J. Savage.
\newblock {Theoretical aspects of quantum electrodynamics in a finite volume
  with periodic boundary conditions}.
\newblock {\em Phys. Rev. D}, 99(3):034510, 2019.

\bibitem{Endres:2015gda}
Michael~G. Endres, Andrea Shindler, Brian~C. Tiburzi, and Andre Walker-Loud.
\newblock {Massive photons: an infrared regularization scheme for lattice
  QCD+QED}.
\newblock {\em Phys. Rev. Lett.}, 117(7):072002, 2016.

\bibitem{Fuda:1973zz}
Michael~G. Fuda and James~S. Whiting.
\newblock {Generalization of the Jost Function and Its Application to Off-Shell
  Scattering}.
\newblock {\em Phys. Rev. C}, 8:1255--1261, 1973.

\end{thebibliography}

\end{document}